\documentclass[%
 aip,
 amsmath,amssymb,
 reprint,%
]{revtex4-1}

\usepackage{graphicx}
\usepackage{dcolumn}
\usepackage{bm}

\usepackage[utf8]{inputenc}
\usepackage[T1]{fontenc}
\usepackage{mathptmx}
\usepackage{etoolbox}

\usepackage{float}

\usepackage{xcolor}
\usepackage{listings}
\definecolor{codegreen}{rgb}{0,0.6,0}
\definecolor{codegray}{rgb}{0.5,0.5,0.5}
\definecolor{codepurple}{rgb}{0.58,0,0.82}
\definecolor{backcolour}{rgb}{0.95,0.95,0.92}
\lstdefinestyle{mystyle}{
  backgroundcolor=\color{backcolour},   commentstyle=\color{codegreen},
  keywordstyle=\color{magenta},
  numberstyle=\tiny\color{codegray},
  stringstyle=\color{codepurple},
  basicstyle=\ttfamily\footnotesize,
  breakatwhitespace=false,         
  breaklines=true,                 
  captionpos=b,                    
  keepspaces=true,                 
  numbers=left,                    
  numbersep=5pt,                  
  showspaces=false,                
  showstringspaces=false,
  showtabs=false,                  
  tabsize=2
}
\makeatletter
\def\@email#1#2{%
 \endgroup
 \patchcmd{\titleblock@produce}
  {\frontmatter@RRAPformat}
  {\frontmatter@RRAPformat{\produce@RRAP{*#1\href{mailto:#2}{#2}}}\frontmatter@RRAPformat}
  {}{}
}%
\makeatother
\begin{document}

\preprint{AIP/123-QED}

\title[Software for Chemical Interaction Networks]{SCINE --- Software for Chemical Interaction Networks}

\author{Thomas Weymuth}
 \affiliation{ETH Zurich, Department of Chemistry and Applied Biosciences, Vladimir-Prelog-Weg 2, 8093 Zurich, Switzerland}
\author{Jan P.~Unsleber}
 \affiliation{ETH Zurich, Department of Chemistry and Applied Biosciences, Vladimir-Prelog-Weg 2, 8093 Zurich, Switzerland}
\author{Paul L.~T\"urtscher}
 \affiliation{ETH Zurich, Department of Chemistry and Applied Biosciences, Vladimir-Prelog-Weg 2, 8093 Zurich, Switzerland}
\author{Miguel Steiner}
 \affiliation{ETH Zurich, Department of Chemistry and Applied Biosciences, Vladimir-Prelog-Weg 2, 8093 Zurich, Switzerland}
 \author{Jan-Grimo Sobez}
 \affiliation{ETH Zurich, Department of Chemistry and Applied Biosciences, Vladimir-Prelog-Weg 2, 8093 Zurich, Switzerland}
\author{Charlotte H. M\"uller}
 \affiliation{ETH Zurich, Department of Chemistry and Applied Biosciences, Vladimir-Prelog-Weg 2, 8093 Zurich, Switzerland} 
\author{Maximilian M\"orchen}
 \affiliation{ETH Zurich, Department of Chemistry and Applied Biosciences, Vladimir-Prelog-Weg 2, 8093 Zurich, Switzerland}
\author{Veronika Klasovita}
 \affiliation{ETH Zurich, Department of Chemistry and Applied Biosciences, Vladimir-Prelog-Weg 2, 8093 Zurich, Switzerland}
\author{Stephanie A. Grimmel}
 \affiliation{ETH Zurich, Department of Chemistry and Applied Biosciences, Vladimir-Prelog-Weg 2, 8093 Zurich, Switzerland}
\author{Marco Eckhoff}
 \affiliation{ETH Zurich, Department of Chemistry and Applied Biosciences, Vladimir-Prelog-Weg 2, 8093 Zurich, Switzerland}
 \author{Katja-Sophia Csizi}
 \affiliation{ETH Zurich, Department of Chemistry and Applied Biosciences, Vladimir-Prelog-Weg 2, 8093 Zurich, Switzerland}
 \author{Francesco Bosia}
 \affiliation{ETH Zurich, Department of Chemistry and Applied Biosciences, Vladimir-Prelog-Weg 2, 8093 Zurich, Switzerland}
\author{Moritz Bensberg}
 \affiliation{ETH Zurich, Department of Chemistry and Applied Biosciences, Vladimir-Prelog-Weg 2, 8093 Zurich, Switzerland}
\author{Markus Reiher}
 \email{mreiher@ethz.ch (corresponding author)}
 \affiliation{ETH Zurich, Department of Chemistry and Applied Biosciences, Vladimir-Prelog-Weg 2, 8093 Zurich, Switzerland}

\date{April 22, 2024}

\begin{abstract}
 The software for chemical interaction networks (SCINE) project
 aims at pushing the frontier of quantum chemical calculations on molecular structures to a new level. While calculations on individual structures as well as on simple relations between them have become routine in chemistry, new developments have
 pushed the frontier in the field to high-throughput calculations.  Chemical relations may be created by a search for specific molecular properties in a molecular design attempt or they can be defined by a set of elementary reaction steps that form a chemical reaction network. The software modules of SCINE have been designed to facilitate such studies. The features of the modules are 
 (i) general applicability of the applied methodologies ranging from electronic structure (no restriction to specific elements of the periodic table) to microkinetic modeling (with little restrictions on molecularity),
 full modularity so that SCINE modules can also be applied as stand-alone programs or be exchanged for external software packages that fulfill a similar purpose (to increase options for computational campaigns and to provide alternatives in case of tasks that are hard or impossible to accomplish with certain programs), 
 (ii) high stability and autonomous operations so that control and steering by an operator is as easy as possible,
 and (iii) easy embedding into complex heterogeneous environments for molecular structures taken individually or in the context of a reaction network. A graphical user interface unites all modules and ensures interoperability. All components of the software have been made available
 open source and free of charge.
\end{abstract}

\maketitle

\section{Introduction}
\label{sec:introduction}

The past decades have seen a steady rise in computing power. Vast computational resources have become available, allowing for large-scale computing campaigns such as virtual high-throughput screening\cite{Shoichet2004, Pyzer-Knapp2015}. Hence, more challenging tasks have become possible of which the exploration of complex chemical reaction networks (CRNs) is one example\cite{Dewyer2018, Vazquez2018, Simm2019, Maeda2021, Baiardi2022}.
However, the complexity of these tasks implies that their automation is not straightforward. In fact, while standard quantum chemical software provides building blocks, these need to be well integrated into a meta-algorithm that orchestrates and steers the computational campaign to allow for the automated execution of complex exploration procedures with only little human input.
Hence, a new kind of software for such a purpose is needed, which our software for chemical interaction networks (SCINE) project aims to provide. Under the umbrella of SCINE, a large number of different software packages is being developed. The primary focus of SCINE, uniting all these modules, is to provide software for predictive quantum chemistry.

\begin{figure}[htb!]
	\centering
	\includegraphics[width=1.0\columnwidth]{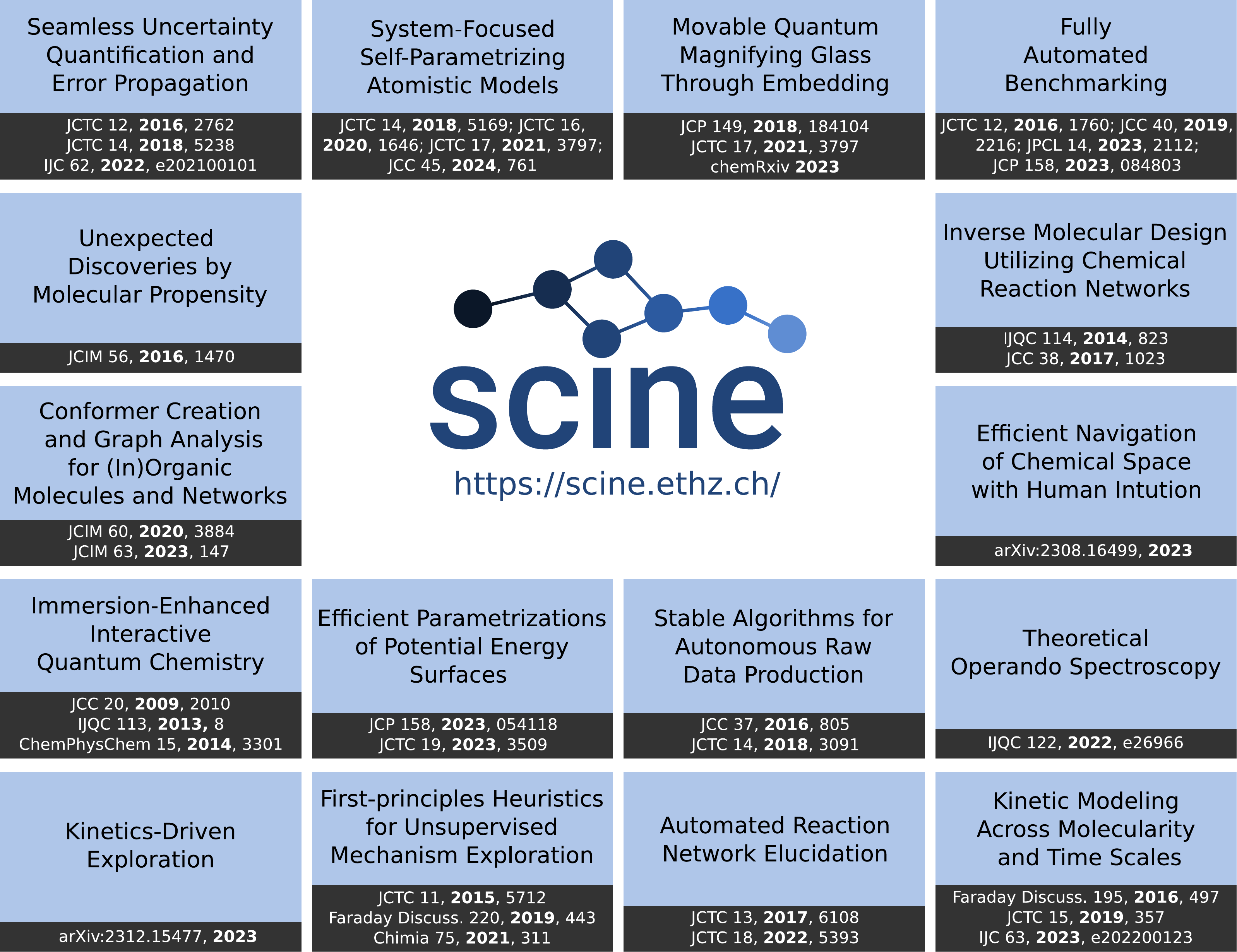}
	\caption{\label{fig:concepts} \small Central scientific concepts implemented by the SCINE software modules with corresponding references (in the black boxes).}
\end{figure}

The software modules created within the SCINE framework revolve around a common set of concepts and ideas shown in Fig.~\ref{fig:concepts}. These ideas have been developed in our research group since 2008. The first concept was interactive and haptic quantum chemistry\cite{Marti2009, Haag2013, Haag2014a}, 
which allows a human to manipulate a chemical system and experience its response in real time 
that can be exploited in
the exploration of chemical reaction space\cite{Haag2011}. 
The ultrafast calculations that underlie interactive quantum mechanical studies led us to introduce the concept of molecular propensity\cite{Vaucher2016b} for assessing the likelihood of a reactive system to change a currently explored Born--Oppenheimer surface (e.g., through reduction, oxidation, spin change, and protonation) in order to screen for unexpected reactive events that allow for new discoveries during mechanistic explorations.
The general structure of how we explore chemical reaction space has already been sketched in Ref.~\citenum{Haag2014} (see Fig.~1 in that paper).
Uncertainty quantification\cite{Mortensen2005, Simm2017, Pernot2022, Weymuth2024}
is an important concept, which we introduced in our quantum chemical reaction exploration framework in order to increase its predictive power\cite{Simm2016, Simm2018, Reiher2022}. 
In order to make automated and interactive quantum chemical studies as reliable as possible, we have been working on stable algorithms to facilitate autonomous raw data production\cite{Vaucher2016, Muhlbach2016, Vaucher2018},
since human intervention in case of troubled or failed calculation would not be possible in such settings.
To increase the efficieny of unsupervised automated reaction network explorations, we introduced first-principles heuristics to screen for reactive events in a way that is agnostic with respect to element types as it is based on the interpretation of the electronic wave function of the reactants\cite{Bergeler2015, Grimmel2019, Grimmel2021}. When brought together, these at first sight somewhat disparate concepts have provided the foundation for the fully automated exploration of complex chemical reaction networks that have now become possible within our SCINE framework. In 2017, we presented the first version of our Chemoton software implementing such explorations\cite{Simm2017a}, which was the successor of our initial implementation of automated reaction exploration algorithms utilizing first-principles heuristics\cite{Bergeler2015}. 
Starting in 2018, we have begun a major rewrite of the entire software base, implementing a flexible and modular architecture. This resulted in the SCINE framework as it is available today and presented in this work. This flexibility allows one to grow the SCINE framework and incorporate new ideas. The latest additions are the ability to harness human intuition for the efficient navigation in chemical space with a steering wheel\cite{Steiner2024}, microkinetical modeling to efficiently drive explorations forward\cite{Bensberg2023a}, and the incorporation of the new concept of lifelong machine-learning models\cite{Eckhoff2023}.

The first-principles prediction of reaction mechanisms in chemistry encompasses the ability to find all possible reactions between a set of reactants, no matter what these reactants are. Therefore, one has to be able to cope with complicated electronic structures (such as molecules with multiconfigurational character), very large molecules
or molecular aggregates (such as molecules adsorbed on a surface), and solvation effects, 
to name only a few challenges. As a consequence, a software stack allowing to study a chemical system from a truly holistic point of view needs to provide broad and stable functionality. This, in turn, results in challenges concerning the flexibility, ease of use, and long-term maintainability of the software.

Within SCINE, we address these challenges by a strictly modular architecture, a loose coupling between the individual modules, and consistent interfaces between them. Individual modules which are only loosely coupled in terms of their source code (\textit{i.e.}, the flow of data between them is clearly defined and limited to only a few instances) are easier to maintain than a complex monolithic architecture. At the same time, this makes developing the software easier and less error-prone, as so-called side effects (\textit{i.e.}, unexpected and unwanted effects upon changing the source code) are largely minimized. Finally, this approach naturally leads to individual modules that are fully functional programs in their own right with useful features. For less complex tasks, it is therefore sufficient to install only a single or a few required modules without having to deal with a particularly complicated software stack. (We note, however, that also the entire SCINE toolchain is straightforward to install by relying on the so-called bootstrap procedure offered by the module Puffin; see below for more details.)

Complex computational procedures will require many SCINE modules to work together. Well-defined interfaces between these modules largely simplify the creation of workflows, leveraging the functionality provided by different modules. We strive to make these interfaces as well as the overall architecture as similar and consistent as possible across the entire SCINE software stack. Therefore, different modules provide a similar appearance in such a way that someone experienced with the code of a particular module should quickly be able to familiarize themselves with any other module. This standardization of the code is, therefore, of crucial importance to reduce the complexity of working with it.

Finally, we believe that the entire functionality provided by SCINE should be freely available to the academic community. Therefore, we provide all the modules under the permissive three-clause BSD license.

\begin{figure}[htb!]
	\centering
	\includegraphics[width=1.0\columnwidth]{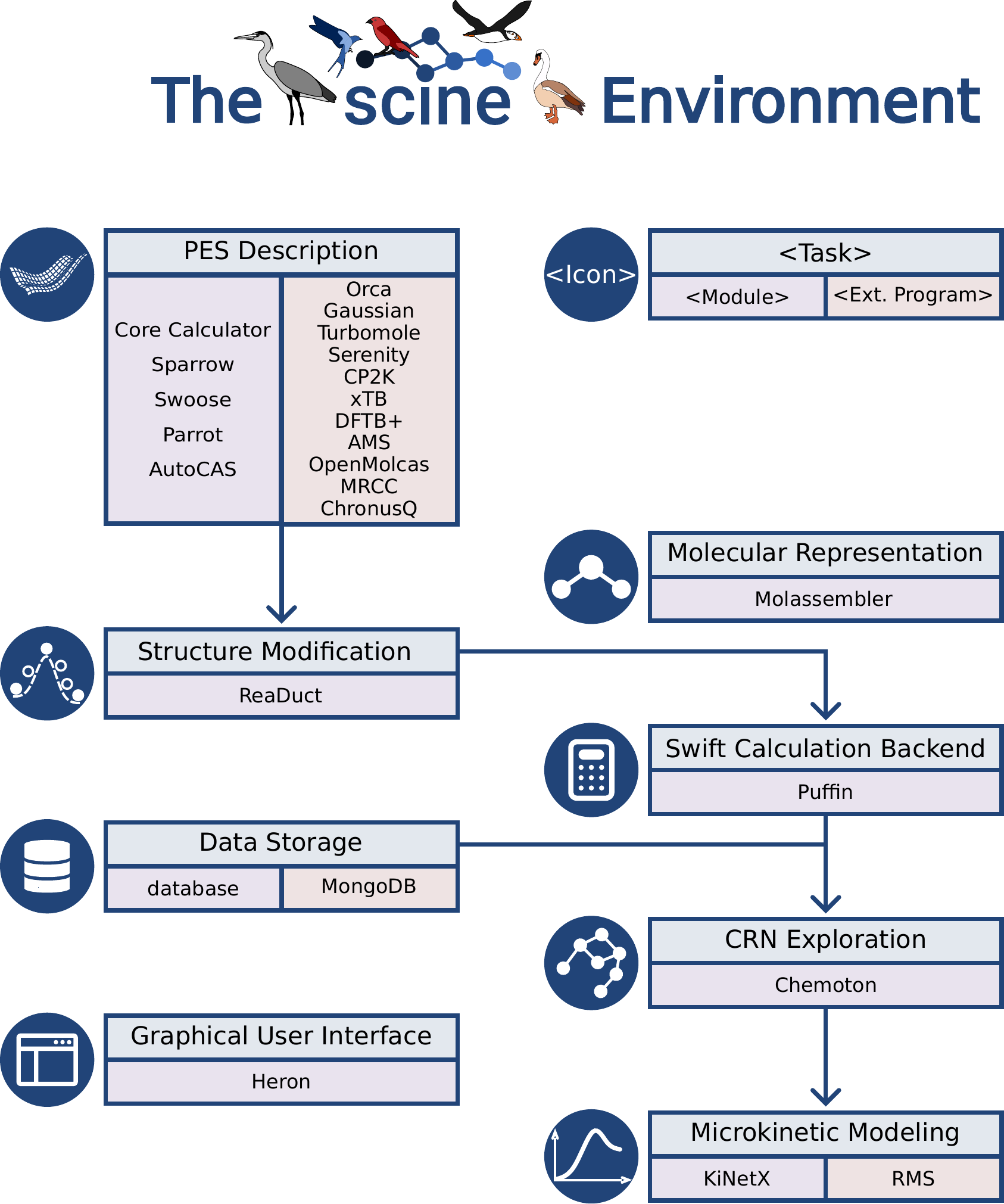}
	\caption{\label{fig:overview} \small An overview of the SCINE modules arranged according to task classes.}
\end{figure}

Fig.~\ref{fig:overview} presents a high-level overview of the SCINE modules presented in this work. In this figure, the modules are grouped according to the tasks they are responsible for. Many modules (for example, Sparrow, Swoose, and Parrot; also many external programs such as ORCA and Turbomole are supported) deliver information on potential energy surfaces (PES), \textit{i.e.}, they calculate electronic energies, nuclear gradients, Hessian matrices, and so on. These quantities are essential inputs for ReaDuct, which is devoted to structure manipulation tasks such as transition state searches. Even though it can be operated in a stand-alone manner, in the context of a high-throughput screening or exploration of a chemical reaction network, a large number of calculations needs to be carried out. For this, we have the calculation backend Puffin. Puffin relies on a database to store all calculation inputs and outputs. Chemoton makes use of these data to drive the exploration of CRNs. Once a network has been established, the module KiNetX can establish a microkinetic model to analyze the concentration fluxes through this network (such a microkinetic analysis can even be applied during a running calculation; see below). Many SCINE modules operate on a graph-based representation of molecular structures; for this we developed the module Molassembler. Finally, there is a graphical user interface (GUI), called Heron, which eases interacting with SCINE modules.

This work is structured as follows: In section~\ref{sec:overview}, we will first give an in-depth overview of important SCINE modules and the data flows between them. Afterwards, every module will be presented in more detail in its own subsection. In these subsections, we will also highlight similar software developments by others whenever appropriate.

\section{Overview of SCINE Modules}
\label{sec:overview}

\begin{figure}[htb!]
	\centering
	\includegraphics[width=1.0\columnwidth]{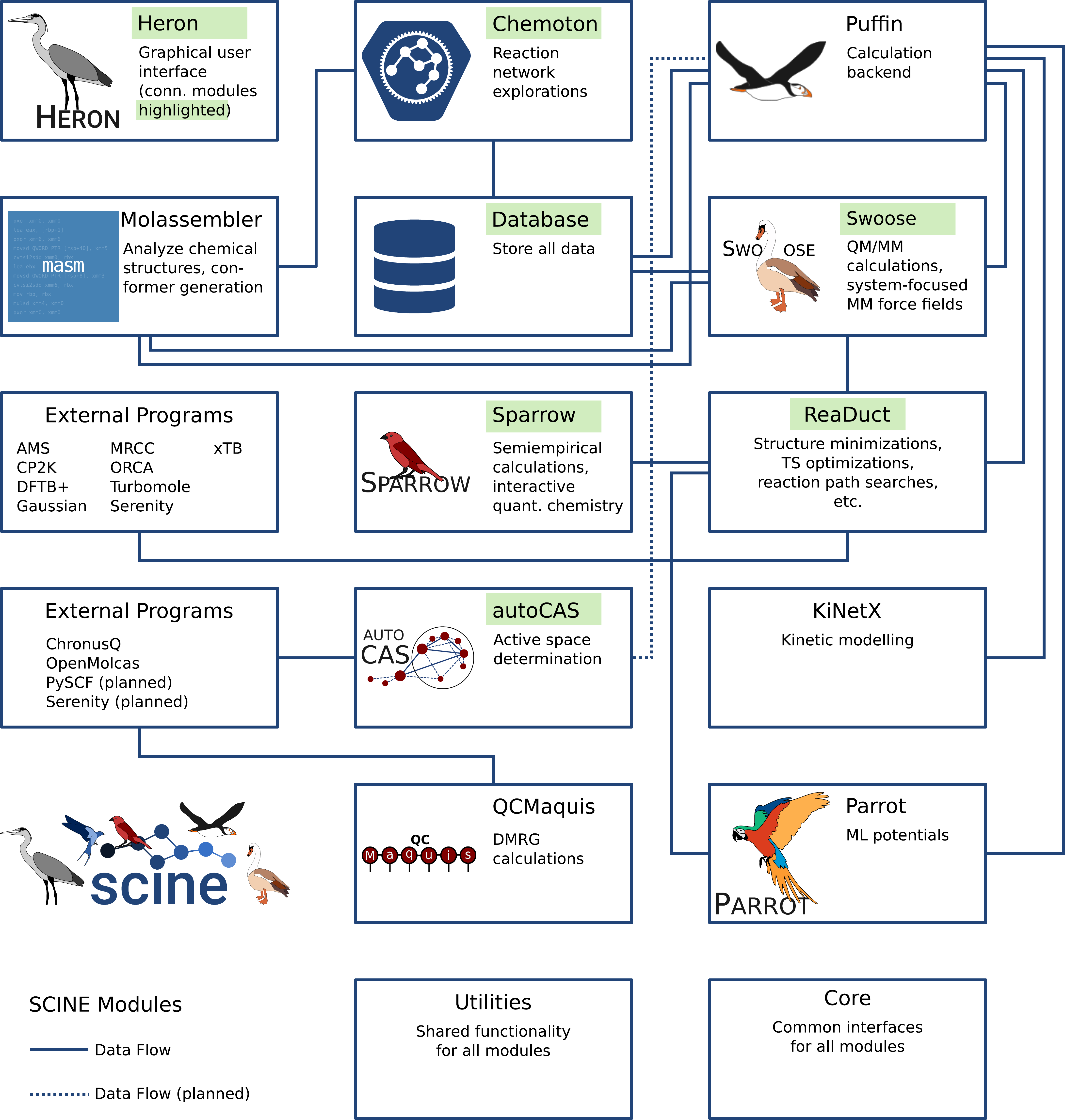}
	\caption{\label{fig:scine} \small SCINE modules and their communication. Data flows from and to Heron, Utilities, and Core are not shown for clarity.}
\end{figure}

A high-level overview of most SCINE modules and of the data flows connecting them is shown in Fig.~\ref{fig:scine}.

Chemoton\cite{chemoton310, Unsleber2022, Simm2017a} is the central module for the exploration of chemical reaction networks. During an exploration, it keeps track of all compounds and reactions found so far (for this bookkeeping, it relies on functionality for the analysis of chemical structures provided by Molassembler\cite{molassembler201, Sobez2020}, see below), properly stores all these data in a database, and sets up new calculations to drive the exploration forward. Chemoton does not carry out these calculations but writes all necessary input data to the database.

The calculations are then carried out by another module called Puffin\cite{puffin130}. Puffin retrieves information on new calculations from the database, carries them out, and then writes the results back to the database. For carrying out a calculation, Puffin typically invokes other SCINE modules. For example, it can call Swoose\cite{swoose100, Brunken2020, Brunken2021} for QM/MM calculations or KiNetX\cite{kinetx200, Proppe2019, Bensberg2023} for microkinetic modeling. For most calculations, however, Puffin relies on ReaDuct\cite{readuct510}. The ReaDuct module carries out a wide range of structure-manipulating tasks such as structure minimizations, transition state searches, intrinsic reaction coordinate calculations, B-spline interpolation and optimization of reaction paths\cite{Vaucher2018}, and Newton trajectory calculations\cite{Unsleber2022}.

For these calculations, ReaDuct in turn relies on quantum chemistry software, which delivers properties such as electronic energies, nuclear gradients, and Hessian matrices. ReaDuct interfaces to many external program packages. AMS\cite{teVelde2001}, CP2K\cite{Kuhne2020}, DFTB+\cite{Hourahine2020}, Gaussian\cite{gaussian09}, MRCC\cite{Kallay2020}, ORCA\cite{Neese2022}, Turbomole\cite{Franzke2023}, Serenity\cite{Unsleber2018, Niemeyer2022}, and xTB\cite{Bannwarth2021} are currently supported.

Apart from these external software packages, ReaDuct is also interfaced to Sparrow\cite{sparrow500, Husch2018a, Bosia2022}, which is a SCINE module providing many semi-empirical methods, such as AM1, PM6, DFTB3, and also the OMx methods\cite{Bosia2023}. Moreover, Sparrow is the backend for interactive quantum chemical calculations (see below).

Molassembler is a general-purpose cheminformatics toolkit for the modeling of organic {\it and} inorganic molecules\cite{Sobez2020}. It represents a molecule as an annotated graph which includes information about the local coordination geometries of the atoms and stereochemical information. During an exploration, it is a key ingredient to compare structures. Since Molassembler can also create three-dimensional coordinates from its graph representation, it can be used to create conformers of a given molecule.

Large-scale explorations typically require millions of quantum chemical calculations. Therefore, fast methods with sufficient accuracy are of vital importance. Machine learning potentials can offer both high accuracy and high speed when trained on sufficiently accurate data. However, such potentials require extensive and costly training and often have difficulties learning additional data without ``forgetting'' previous knowledge. Moreover, most structural descriptors cannot efficiently represent many different atom types. We recently addressed these problems by introducing the new concept of lifelong machine learning potentials\cite{Eckhoff2023} implemented in the SCINE module Parrot. Other well-known machine learning potentials such as ANI\cite{Smith2017, Devereux2020}, M3GNet\cite{Chen2022}, and MACE\cite{Batatia2022} are also available in Parrot.

Another feature which is critical for large-scale quantum chemical calculations is parallelization. Performance-critical parts of the SCINE framework are parallelized with OpenMP. Furthermore, the various parallelization features of the external quantum chemistry packages mentioned above are also supported: When invoking ReaDuct, the user can specify the number of cores which should be utilized. This setting is then forwarded to the underlying software. For example, in the case of ORCA, the number of cores will be written to the corresponding ORCA input file, while in the case of Turbomole, the environment variable \texttt{PARNODES} will be set.

Since Chemoton is based on the first principles of quantum mechanics, it is applicable to any chemical system. In practice, this might require specific quantum chemical methods. For example, for very large systems such as proteins, even fast semi-empirical methods like the ones provided by Sparrow are typically too time-intensive for large-scale explorations. In this case, one can resort to molecular mechanics calculations. To enable such type of modeling out of the box, the SCINE software stack includes a module called Swoose. Swoose can provide energies and forces from molecular mechanics for any chemical system and run molecular mechanics simulations. If no parameters are available for the system, it can automatically parametrize a system-focused force field from quantum chemical reference calculations\cite{Brunken2020}. Large systems for which accurate reference calculations would be too costly are automatically fragmented. Moreover, Swoose is also able to carry out QM/MM calculations, in which a key region within a system is automatically defined and then described by an electronic structure model, while the rest of the system is described by a standard force field (supported by machine learning corrections)\cite{Brunken2021}.

Open-shell systems often require a multi-configurational wavefunction ansatz to be described qualitatively correctly. Multi-configurational methods such as the density matrix renormalization group (DMRG)\cite{White92, White93, Chan2008, Zgid2009, Marti2010, Schollwoeck2011, Chan2011, Wouters2013, Kurashige2014, Olivares2015, Szalay2015, Yanai2015, Baiardi2020} typically require the definition of a so-called active orbital space on input. However, this active space is often not trivial to determine. Our autoCAS algorithm\cite{Stein2016a, Stein2016, Stein2017a}, implemented in a module of the same name\cite{Stein2019, autocas210}, provides an automated way to reliably determine the active space of any molecule. Moreover, once the active space is determined, autoCAS can also invoke software to carry out multi-configurational calculations. Besides QCMaquis\cite{Keller2015}, which is our software for DMRG calculations and driven by autoCAS via OpenMolcas\cite{LiManni2023}, autoCAS is also interfaced to ChronusQ\cite{Williams-Young2020}, PySCF\cite{Sun2020}, and Serenity\cite{Unsleber2018}. We note that QCMaquis can do much more than just calculate the ground state electronic energy of a molecule. It can also optimize excited states\cite{Baiardi2019, Baiardi2022a}, solve the vibrational Schr\"odinger equation for anharmonic potential energy surfaces\cite{Baiardi2017, Glaser2023},
and carry out quantum dynamics simulations\cite{Baiardi2019a, Baiardi2021}. Furthermore, it features embedding approaches\cite{Dresselhaus2015, Hedegard2016}, a non-orthogonal state interaction method\cite{Knecht2016}, relativistic calculations\cite{Battaglia2018}, multi-component DMRG\cite{Muolo2020, Feldmann2022}, and N-electron valence state perturbation theory calculations\cite{Freitag2017}.

A central module is Heron\cite{heron100}, our graphical user interface. It aims to make the entire SCINE software ecosystem easily accessible. Heron provides the ability to browse the contents of a database, visualize and analyze existing reaction networks, and operate Chemoton (for example, with the so-called Steering Wheel\cite{Steiner2024}). Moreover, it is a graphical user interface for Swoose, ReaDuct, autoCAS, and Sparrow. Together with the latter, it provides the ability to interactively
carry out quantum chemical calculations, in which the response of a molecule to structural manipulations can be experienced in real time. This can even be coupled with haptic devices, allowing a person to literally feel the forces acting on atoms in a molecule\cite{Marti2009, Haag2013, Haag2014a, Weymuth2021}.

For the sake of completeness, we mention two modules that are essential for the entire SCINE software stack, but they are typically not very visible when carrying out calculations. The first of these modules is Core\cite{core600}, which provides the common so-called Calculator interface. This interface is crucially important for the high interconnectivity yet loose coupling of the individual SCINE modules. The second module is called Utilities\cite{utilities900}; it contains much functionality needed by many of the other modules. For example, it contains optimization algorithms, data structures for objects such as atomic orbitals and density matrices, functions to read and write molecular coordinates to standard file formats such as XYZ and PDB, and so forth.

In the following, we present SCINE modules in more detail. For each module, we describe the field of application, its input and output, as well as its application programming interface (API), and we illustrate the latter with code snippets.

\subsection{Molassembler}
\label{sec:molassembler}

\subsubsection{Field of application}
Molassembler\cite{molassembler201, Sobez2020} provides characterization, manipulation, and conformer generation for organic and inorganic molecules. In principle, it can handle any organic or inorganic molecule, including haptic and multidentate ligands and almost arbitrary coordination spheres. Its general approach to the characterization of  molecules and its broad applicability makes it distinct compared to similar programs such as RDKit\cite{RDKit2023} or openBabel\cite{OBoyle2011}, which focuses on organic molecules, or molSimplify\cite{Ioannidis2016, Edholm2023}, AARON\cite{Guan2018} and DENOPTIM\cite{Foscato2019}, which provide utilities to design new molecules, or CREST\cite{Pracht2020} for conformer generation.

Molassembler characterizes a molecule's connectivity by a graph and the local geometry of each atom by polyhedral shapes. In organic chemistry, atoms typically have only up to four bonding partners, leading at most to a tetrahedral local environment. By contrast, transition metal atoms can have significantly more neighbors and diverse coordination geometries. Therefore, Molassembler supports polyhedra for the local geometry up to icosahedra and cuboctahedra and can be applied to the full spectrum of organic and inorganic chemistry. Furthermore, it supports multidentate and haptic ligands. Molassembler differentiates stereoisomers through a ranking algorithm based on generalizing the ranking rules for organic molecules provided by the International Union of Pure and Applied Chemistry\cite{Favre2013}. The abstract molecule representation provided by Molassembler can be manipulated to change the connectivity in the molecule and replace or edit ligands to generate new molecules while retaining any chiral information in the remaining molecule wherever possible. Furthermore, Molassembler can generate conformers (either by by enumerating the rotamers of the molecule or by relying on Distance Geometry\cite{Blaney2007} for a stochastic conformer generation), generating initial Cartesian coordinates for the conformer guesses before refining the coordinates. 

\subsubsection{Scientific background}
In the context of SCINE, one task delegated to Molassembler is the interpretation of the results of quantum chemical calculations into molecular graphs. Because the input structures to these calculations may be of unknown stability or of purposeful instability such as in the case of reactivity explorations, no information regarding the bond connectivity or stereochemistry of the input structure may be transferred onto the output structure of the calculation. Due to these limitations, the interpretation of a structure proceeds with only its atom-labeled spatial positions, and possibly a quantum chemically calculated bond order matrix, as inputs.

Prior to the inference of stereochemical information, possibly available bond order information is combined with internuclear distances into connectivity. The resulting graph can be split into its connected components, each of which represent either a molecule or a singular atom.

Molassembler represents atom-centered stereochemistry as a discrete local polyhedral shape combined with a surjection from shape vertices to the chemically ranked ligand binding sites. There are a number of subtleties to this representation, of which it is sufficient here to mention only one: The discretization of varied local coordination geometries into idealized polyhedral shapes necessarily limits the capacity to represent structures other than ground-state minima, since transition states or excited states can exhibit significantly irregular local shapes.

The classification of an arbitrary local coordination geometry or bonding pattern into discrete shapes is settled on a protocol where the distance between input and shape is quantified by a continuous shape measure\cite{Pinsky1998}. The time-determining step of the algorithm involves finding a point to vertex matching that minimizes a quaternion fit. The implementation in Molassembler reduces the factorial complexity of this step by exploiting rotational symmetries of the target shape and a heuristic considering a rotational fit sufficiently converged with five vertex matchings. This allows for calculations with shapes up to twelve vertices in size on the order of seconds. Additionally, the origin of the spherical shapes is pre-matched to the atomic nucleus at the center of the local geometry, adding more information into the quaternion fit. A complication stemming from the incomparability of continuous shape measures with respect to different target shapes has been solved by further calculating the probability that each measure is the result of a random input geometry. The shape whose continuous shape measure is least likely to have been calculated from a random input geometry is selected. We refer to the original publication in Ref.~\citenum{Sobez2020} for further details on the algorithms implemented.

From the point to vertex matching gained from selecting a local shape and chemically ranking ligand and substituent binding sites, local stereoinformation can be constructed.

\subsubsection{API}
Molassembler can be used through its Python API or as a C++ library. It generates its abstract molecule representation based on either Cartesian coordinates and bonding pattern, its SMILES\cite{Weininger1988} or its InChi\cite{Heller2015} representation. Furthermore, it can read and write its own molecule representation as a JSON document or binary string.

\begin{figure*}
    \centering
    \includegraphics[width=\textwidth]{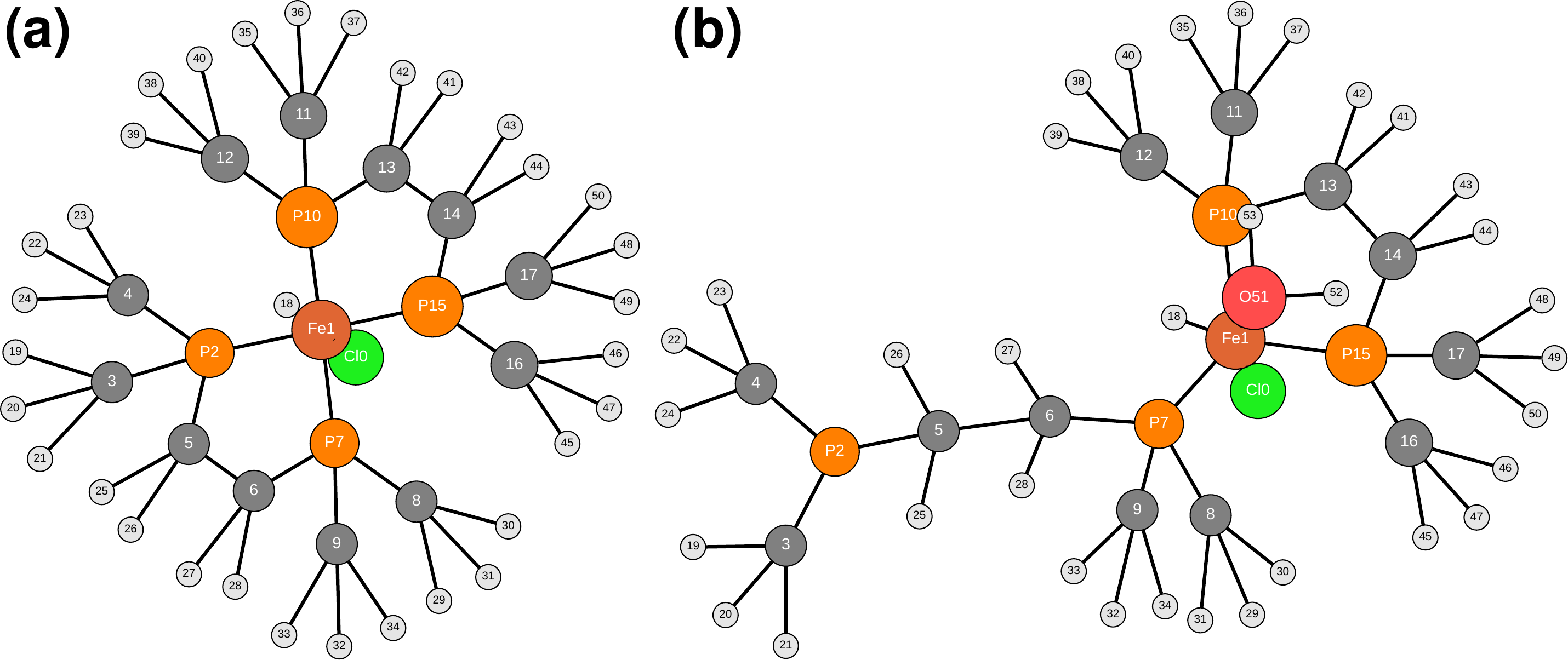}
    \caption{Example iron--phosphorus complex manipulated with Molassembler.}
    \label{fig:molassembler-examples}
\end{figure*}

We demonstrate a simple molecule manipulation with Molassembler in the code extract~\ref{c:molassembler-code}. We first create Molassembler's internal molecule representation of an iron--phosphorus complex by reading its SMILES string in line 5. Then, we create the vector graphic in Fig.~\ref{fig:molassembler-examples}(a), displaying the molecule, cutting one of the iron-phosphorus bonds in line 8, and adding a water ligand in lines 10 to 12. Finally, we save a vector graphic displaying the final molecule [see Fig.~\ref{fig:molassembler-examples}(b)].

\begin{lstlisting}[language=Python, caption=Example code for molecule manipulations with Molassembler., label=c:molassembler-code]
import scine_utilities as utils
from scine_molassembler import io

complex_smiles = "Cl[FeH-4]56([P+](C)(C)CC[P+]5(C)C)[P+](C)(C)CC[P+]6(C)C"
fe_complex = io.experimental.from_smiles(complex_smiles)
io.write("mol.svg", fe_complex)

fe_complex.remove_bond(1, 2)

new_atom_index = fe_complex.add_atom(utils.ElementType.O, 1)
fe_complex.add_atom(utils.ElementType.H, new_atom_index)
fe_complex.add_atom(utils.ElementType.H, new_atom_index)

io.write("mol2.svg", fe_complex)                               
\end{lstlisting}

Based on Molassembler's molecule representation, we can then generate conformers of the molecule, as shown in the code extract~\ref{c:molassembler-conformer}. In this example, we let Molassembler generate an ensemble of 100 conformers.

\begin{lstlisting}[language=Python, caption=Conformer generation with Molassembler., label=c:molassembler-conformer]
from scine_molassembler import dg
results = dg.generate_ensemble(fe_complex, 100, seed=42)
\end{lstlisting}
The conformer generation's result (\texttt{results}) is a list containing the Cartesian coordinates of the conformers, which could be analyzed further with other tools within the SCINE framework.

\subsection{Utilities}
\label{sec:utilities}

\subsubsection{Field of application}
Certain functionality (\textit{e.g.}, reading in molecular structures from XYZ files) is required in several SCINE modules. Having separate functions for this in every individual module would lead to code which is difficult to maintain. Therefore, we created the separate module SCINE Utilities\cite{utilities900} to collect functionality shared across the entire SCINE project. SCINE Utilities is rarely used in a stand-alone manner (even though it could) since the functionality provided by it is typically needed in the more complex tasks and workflows provided by other SCINE modules.

From the wealth of functionality provided by Utilities we shall explicitly mention here data structures to store objects such as molecular structures, atomic orbitals, and density matrices; interfaces to external quantum chemistry programs such as Turbomole\cite{Franzke2023} and ORCA\cite{Neese2020}; different optimization algorithms (for example, the well-known L-BFGS algorithm\cite{Nocedal1980}); a self-consistent field algorithm including orbital steering\cite{Vaucher2017}; code for calculating thermodynamic quantities such as the Gibbs free energy relying on the rigid-rotor--harmonic-oscillator model; an implementation of Grimme's semiclassical D3 dispersion corrections\cite{Grimme2010}; basic molecular dynamics simulation algorithms; functions to handle different chemical file formats (like XYZ and PDB); code for the placement of explicit solvent molecules around a solute; and finally, a lot of technical functionality required across all SCINE modules (for example, data structures and functions to handle calculation settings).

\subsubsection{API}
The easiest way to interact with the Utilities module is via its Python bindings. As an example, the code in Listing~\ref{code:python_utilities} first uses data structures provided by Utilities to manually construct a helium dimer with an interatomic distance of 5.0\,bohr. Then, this structure is optimized (with the BFGS algorithm as implemented in the Utilities module) by minimizing its energy; the energy is obtained from a simple Lennard-Jones potential, which is also provided by the Utilities module. Finally, the optimized structure is written to an XYZ file.

\begin{lstlisting}[language=Python, label=code:python_utilities, caption=Simple structure minimization with the Utilities module.]
import scine_utilities as utils

elements = [utils.ElementType.He, utils.ElementType.He]
positions = [[0.0, 0.0, 0.0], [5.0, 0.0, 0.0]]
structure = utils.AtomCollection(elements, positions)

calculator = utils.core.get_calculator("lennardjones")
calculator.structure = structure
calculator.set_required_properties([utils.Property.Gradients, utils.Property.Energy])

opt_log = utils.core.Log()
minimized_structure = utils.geometry_optimize(calculator, opt_log, utils.Optimizer.Bfgs)

utils.io.write("minimized_structure.xyz", minimized_structure)
\end{lstlisting}

\subsection{Calculator Interface}

The core of the modular SCINE infrastructure is a custom \texttt{Calculator} interface providing the (abstract) functionality to calculate electronic structure information, such as energy and gradients, for a given set of nuclear coordinates. We have developed interfaces to various programs that allow one to choose methods ranging from molecular mechanics, semi-empirical methods, density functional theory (DFT) to coupled cluster theory. These interfaces will be discussed in more detail in the following sections.

\subsection{Sparrow}

\subsubsection{Field of application}
Sparrow\cite{sparrow500, Bosia2022, Bosia2023} specializes in ultra-fast quantum chemical calculations with semi-empirical methods, which are essential in applications such as interactive quantum chemistry or large-scale reaction network exploration. Calculations with semi-empirical methods based on the neglect of diatomic differential overlap approximation\,---\,NDDO methods\,---\,as well as with methods based on the density functional tight binding approach\,---\,DFTB methods\,---\,are available (see below). Ground state calculations yield electronic energies, Hessian matrices, nuclear gradients, bond orders, and thermodynamic properties such as Gibbs free energies. Electronically excited states can also be calculated\,---\,in a configuration interaction approach with NDDO methods or by solving the linear response eigenvalue problem in a time-dependent approach with DFTB methods.

Through the interactive user interface Heron\cite{heron100}, an operator is able to interact with a molecular system of interest in real time. This is made possible by real-time quantum chemical calculations\cite{Haag2013} implemented as ultrafast semi-empirical approaches in Sparrow.
Specifically, the molecular structure(s) relax toward an energy-minimum structure while the operator may manipulate the structure simultaneously\cite{Haag2014, Vaucher2016, Vaucher2017} (\textit{e.g.}, in search of viable reaction paths\cite{Haag2011}). 
For a smooth operator experience of these manipulations, a high-frequency continuous feedback of electronic energies and gradients of at least 25\,Hz is required\cite{Mark1996, Ruspini1997}.

We emphasize that Heron does not only offer an interactive experience with the computer mouse, but further allows the operator to feel the quantum chemical forces on atomic nuclei in the context of haptic quantum chemistry\cite{Marti2009}. That is, the gradient of the energy at the position of a selected atom can be experienced by an operator's tactile sense as attraction or resistance along the direction of manipulation through a force-feedback haptic device\cite{Marti2009}.

\subsubsection{Scientific background}
Sparrow implements semi-empirical methods of the neglect of the diatomic differential overlap (NDDO) and density functional tight binding (DFTB) families. These allow for ultrafast, albeit approximate quantum chemical calculations. 

Within the NDDO approximation, the Hartree--Fock formalism is modified by neglecting the differential overlap between atomic orbitals $\chi_{\mu^I}$ and $\chi_{\nu^J}$ on different atomic centers, \textit{i.e.}, $\delta_{IJ}\chi_{\mu^I}\chi_{\nu^J} \approx \phi_\mu \phi_\nu$, where $\chi_{\mu^I}$ denotes the basis function $\mu$ of an atomic orbital expansion in a non-orthogonal basis centered on the atom $I$, and $\phi_\mu$ is the corresponding basis function in an orthogonal basis, such that
\begin{equation}
\phi_\nu = \sum_\mu^M \left(S^{-\frac{1}{2}}\right)_{\mu\nu}\chi_\mu \; .
\end{equation} 
As a consequence, the two-electron repulsion integrals in an orthogonalized atomic orbital basis can be approximated in terms of two-electron integrals in a non-orthogonal atomic orbital basis as
\begin{equation}
\langle\phi_{\mu}\phi_{\nu}|\phi_{\lambda}\phi_{\sigma}\rangle \approx \delta_{IJ}\delta_{KL}\langle\chi_{\mu^I}\chi_{\nu^J}|\chi_{\lambda^K}\chi_{\sigma^L}\rangle \; .
\end{equation}
In all NDDO methods, all three- and four-center two-electron repulsion integrals are neglected. The advantage is twofold: First, their computational expensive calculation is avoided, and second, the NDDO Fock matrix ${}^{\textrm{NDDO}}\boldsymbol{F}$
assembled from these integrals is already expressed in an approximately orthogonal basis.
NDDO methods solve the eigenvalue problem
\begin{equation}
  {}^{\textrm{NDDO}}\boldsymbol{F} {}^{\textrm{NDDO}}\boldsymbol{C} = {}^{\textrm{NDDO}}\boldsymbol{\varepsilon}  {}^{\textrm{NDDO}}\boldsymbol{C} \; ,
\end{equation}
where the ordinary Fock, molecular orbital, and diagonal orbital energy matrices are substituted with their NDDO counterparts. The overlap matrix is set to be equal to the identity matrix by virtue of the orthogonalized basis in which the NDDO calculation is carried out. The specifics of the calculation of the Fock matrix are heavily method-dependent and can be found in Ref.~\citenum{Husch2018a}.

By contrast, DFTB methods are constructed by expanding the energy functional with respect to the electronic density to various degrees of a Taylor series expansion of $\rho$ around a a reference density $\rho_0$,
\begin{align}
\begin{split}
    {}^{\textrm{DFTB}}E[\rho_0 + \delta \rho] & = E^{(0)}[\rho_0] + E^{(1)}[\rho_0 + \delta \rho] \\
    & + \frac{1}{2}E^{(2)}[\rho_0 + \left(\delta \rho\right)^2] + \frac{1}{6}E^{(3)}[\rho_0 + \left(\delta \rho\right)^3] \\
    & +... \quad .
\end{split}    
\end{align}
The DFTB0, DFTB2, and DFTB3 models include terms up to the first, second, and third order, respectively.
DFTB0 calculations do not require iterative solutions until self-consistency has been reached; they include the zeroth and first order of the energy functional expansion.
$E^{(1)}[\rho_0 + \delta \rho]$ is obtained by solving the eigenvalue problem
\begin{equation}
\boldsymbol{\mathcal{H}}^{(0)} \boldsymbol{C} = \boldsymbol{S} \boldsymbol{C}\boldsymbol{\varepsilon} \quad,
\end{equation}
with $\boldsymbol{C}$ being the coefficients of the molecular orbitals, $\boldsymbol{\mathcal{H}}^{(0)}$ and $\boldsymbol{S}$ being the tabulated Hamiltonian and overlap matrices, respectively, appropriately rotated according to the Slater--Koster 
rules, and $\boldsymbol{\varepsilon}$ the diagonal matrix of the orbital energies. The energy contribution is therefore equal to 
\begin{equation}
    E^{(1)}[\rho_0 + \delta \rho] = \sum_i^M n_i\varepsilon_{i}\quad ,
\end{equation}
with $M$ being the number of electrons in the system.
$E^{(0)}[\rho_0]$ is approximated as a sum of diatomic repulsion terms between atoms $I$ and $J$, $V_{\textrm{rep}, IJ}$, modeled to correct the sum of the occupied molecular orbital energies with respect to DFT reference calculations. The resulting DFTB0 energy reads
\begin{equation}
    {}^{DFTB0}E[\rho_0 + \delta \rho] = \sum_i^M n_i\varepsilon_{i} + \frac{1}{2}\sum_I \sum_J V_{\textrm{rep}, IJ} \quad .
\end{equation}
DFTB2 includes the effects of a self-consistent redistribution of Mulliken charges to the energy functional and the elements of the Hamiltonian operator. The corresponding energy expression is given by
\begin{align}
    {}^{DFTB2}E^{(2)} &= \sum_I \sum_J \Delta q_I \Delta q_J \gamma_{IJ} \\
    {}^{DFTB2}\mathcal{H}_{\mu^I\nu^J} &= \mathcal{H}^{(0)}_{\mu^I\nu^J} + \frac{S_{\mu^I\nu^J}}{2} \sum_K^N\Delta q_K \left(\gamma_{IK} + \gamma_{JK}\right) \quad ,
\end{align}
where the notation $\mu^I$ indicates the atomic orbital $\mu$ located on the atomic center $I$, $\Delta q_I$ is the Mulliken charge on atom I, $\gamma_{IJ} = \frac{1}{R_{IJ}} - S$, and $S$ is a model parameter.
DFTB3 introduces another term into the energy density functional expansion. The energy correction and the Hamiltonian coefficients are then given by
\begin{align}
\begin{split}
    {}^{DFTB3}E^{(3)} &= \sum_I \sum_J \Delta q^2_I \Delta q_J \Gamma_{IJ} \; , \\
    {}^{DFTB3}\mathcal{H}_{\mu^I\nu^J} &= {}^{DFTB2}\mathcal{H}_{\mu^I\nu^J} + \frac{S_{\mu^I\nu^J}}{3} \sum_K^N\Delta q_K \Gamma_{IJK} \; , \\
    \Gamma_{IJK} &= \left(\Gamma_{IK}\Delta q_I + \Gamma_{JK} \Delta q_J + \frac{\Delta q_K}{2} (\Gamma_{IK} + \Gamma_{JK})\right) \; ,
\end{split}    
\end{align}
where the function $\Gamma_{IJ}$ represents the derivative of $\gamma_{IJ}$ with respect to the atomic charge, and DFTB3 includes in $\gamma_{IJ}$ another correction to improve on the performance for certain systems.    

The semi-empirical methods available in Sparrow provide fast and reliable electronic energies and gradients. 
However, the calculation cost still scales with the number of atoms.
Therefore, to guarantee the required high frequency for seamless visual and haptic feedback, the potential energy surface is approximated for atomic positions between finalized quantum chemical calculation with a Taylor series expansion around the atomic coordinates
of the last finished electronic structure calculation; the resulting harmonic potential is called mediator or surrogate potential\cite{Vaucher2016}. The mediator potential resembles the local Born--Oppenheimer surface and prohibits escape to regions of configuration space for which no electronic structure information has become available yet.

This real-time interaction is of interest to both researchers who would like to explore their reactive system in a more intuitive and interactive way\cite{Csizi2023b} as well as for learners who are in search for an immersive experience of computational chemistry\cite{Mueller2024a}.

\subsubsection{I/O}
Sparrow can be called from the command line. In the simplest case, the binary is called with only one command-line argument specifying the path to an XYZ file containing Cartesian coordinates:
\begin{lstlisting}[language=Bash, label=code:sparrow_run]
sparrow -x molecule.xyz
\end{lstlisting}
Command-line options are available to specify the calculation. For instance, the ground state energy and the Hessian of a doublet water cation with PM6 can be obtained with the command
\begin{lstlisting}[language=Bash, label=code:sparrow_water]
sparrow -x h2o.xyz -c 1 -s 2 -M PM6 -H
\end{lstlisting}

\subsubsection{API}
Sparrow is integrated into various SCINE workflows through its Python bindings. Python bindings from SCINE Utilities are required to instantiate a calculator object in the first step. The code snippet in Listing~\ref{code:python_sparrow_calculator} shows how the calculator object is created by specifying the desired method. In the second step, the atomic structure is read in from a XYZ file and assigned appropriately. Next, desired (non-default) settings in the form of a dictionary object are defined. Lastly, the desired properties are requested.
\begin{lstlisting}[language=Python, label=code:python_sparrow_calculator, caption=Creating and defining the calculator object.]
import scine_utilities as utils
import scine_sparrow as sparrow

calculator = utils.core.get_calculator("pm6",   "sparrow")
atoms = utils.io.read("h2o.xyz")[0]
calculator.structure = atoms
calculator.settings.update({
    "spin_mode": "unrestricted",
    "molecular_charge": 1,
    "spin_multiplicity": 2
})
calculator.set_required_properties([utils.Property.Energy, utils.Property.Hessian])
\end{lstlisting}

Once the calculator object is fully specified, the actual calculation is run with the command in Listing~\ref{code:python_sparrow_calculation}.

\begin{lstlisting}[language=Python, label=code:python_sparrow_calculation, caption=Carrying out the calculation.]
results = calculator.calculate()
\end{lstlisting}

The results can be saved into dedicated files or simply printed to the console.

\subsection{Swoose}

\subsubsection{Field of application}
If a system of interest is too large for a full quantum mechanical description, a quantum--classical hybrid model must be constructed. Such models generally describe a small subsystem quantum mechanically (QM) and the remainder with a molecular mechanical (MM) model which parametrizes inter- and intramolecular interactions based on experimental or calculated reference data. The module SCINE Swoose provides such MM models, the framework to combine its models with any existing electronic structure theory to construct a QM/MM hybrid model, and workflows specialized for nanoscopic systems. Generally, MM models can be difficult to construct due to the structural complexity of nanoscopic systems and the necessity of MM models to define bonds and charges \textit{a priori}. Swoose overcomes this hurdle by providing a general workflow for structure preparation\cite{Csizi2023a} and a system-focused atomistic model (SFAM)\cite{Brunken2020} that can be generated for each system of interest with automated QM reference data generation by fragmenting the system, carrying out numerous structure optimizations in parallel, and fitting force constants with a partial Hessian procedure. The generated structural and MM model can then be leveraged easily in automated procedures\cite{Csizi2023b}. Reaction mechanisms can be studied with hybrid models, for which the QM region can be automatically determined by another algorithm implemented in Swoose\cite{Brunken2021}. It generates multiple candidates for the QM region around an atom alongside multiple large reference systems, calculates the first-order partial derivatives for each nucleus, \textit{i.e.,} the atomic forces, and compares them against those in the reference systems. The algorithm assumes that small errors in atomic forces are a good descriptor together with energies as shown in Ref.~\citenum{Brunken2021}. It then selects the smallest QM system that features errors below a given threshold. Compared to other approaches to determine a QM region\cite{Sumner2013, Liao2013, Kulik2016, Karelina2017, Brandt2023, Brandt2023a}, to determine protonation states\cite{Dolinsky2004, Dolinsky2007, OBoyle2011, Olsson2011, Anandakrishnan2012, MadhaviSastry2013, Riojas2014, Bochevarov2016, Eastman2017, Reis2020}, and to fit a system specific model\cite{Grimme2014, Vanduyfhuys2015, Welsh2019, Behler2021, Friederich2021, Unke2021, Deringer2021, Musil2021}, Swoose exploits the straightforward parallelization of multiple calculations within the SCINE framework and provides a complete pipeline from experimental data to systematic reaction explorations with QM/MM hybrid models, including real-time calculations\cite{Csizi2023b}. 

\subsubsection{Scientific background}
\label{subsubsec:swoose_scientific_background}

Swoose provides two molecular mechanics models, GAFF\cite{Wang2004} and our system-focused atomistic model SFAM. The latter adopts the following (standard) functional form:
\begin{align}
\begin{split}
        E_{\mathrm{MM}} &= E_{\mathrm{cov}} + E_{\mathrm{noncov}} = E_{r} + E_{{\alpha}}  + E_{{\psi}} + E_{{\phi}} + E_{\mathrm{estat}} \\
        &+ E_{\mathrm{disp}} + E_{\mathrm{rep}} + E_{\mathrm{hb}}.
        \label{eq:sfam}
\end{split}
\end{align}

The potential energy contributions from bonds ($E_{r}$) and angles ($E_{{\alpha}}$) are taken to be harmonic potentials. The dihedral angle terms ($E_{{\psi}}$) are modeled by cosine functions. To describe accurately out-of-plane torsions of planar fragments and to account for inversion barriers of trigonal planar fragments, SFAM incorporates an improper dihedral term ($E_{{\phi}}$), which takes a harmonic form for small out-of-plane angles and a double-well potential for large equilibrium angles. The electrostatic term ($E_{\mathrm{estat}}$) is defined as a pairwise sum of Coulomb interactions between atom-centered partial charges. For repulsive ($ E_{\mathrm{rep}}$) and dispersive ($E_{\mathrm{disp}}$) interactions, we adopt the functional form of Grimme's QMDFF force field\cite{Grimme2014}, which consists of an exponential term for repulsions alongside the D3 semiclassical dispersion corrections\cite{Grimme2010}. Also the SFAM-hydrogen bonding potentials ($E_{\mathrm{hb}}$) for atom triples (hydrogen donor, hydrogen acceptor, hydrogen atom) are inspired by QMDFF.

The amount of required reference data to derive all parameters in Eq.~(\ref{eq:sfam}) for any given structure is rather limited and only comprises the equilibrium structure of the molecule, the corresponding Hessian matrix, the atomic partial charges, and the covalent bond orders. Divide-and-conquer fragmentation schemes allow us to provide scalable variants of the parametrization routine applicable to the nanoscopic scale. For details on the functional form and parametrization procedures, we refer to Ref.~\citenum{Brunken2020}.

After successful parametrization, SFAM can be employed as a classical force field in hybrid QM/MM calculations. As SFAM parameters are always generated for the entire molecular system, the QM region can be dynamically redefined at any time. Furthermore, SFAM can be locally reparametrized if the bonding pattern in the QM region changes through a chemical reaction, but the corresponding atoms are subsequently transferred to the MM region.
In the QM/SFAM model, both mechanical embedding (ME) and electrostatic embedding (EE) are supported. Although the quantum mechanical treatment of interaction energies in the EE scheme is more accurate, the applicability of EE depends on the chosen QM model and on whether the QM Hamiltonian can be equipped with one-electron terms for the charged MM atoms. Within SCINE, xTB and the external programs ORCA and Turbomole can be employed for EE-QMM/MM calculations. For ME, the QM/SFAM energy expression of the nanoscopic system composed of an environment region $\mathcal{E}$ and a quantum region $\mathcal{Q}$ is defined as
\begin{equation}
        E_\mathrm{QM/SFAM}^\mathrm{ME} = E_{\mathrm{QM, ME}}^\mathcal{Q} + E_\mathrm{SFAM}^{\mathcal{Q}+\mathcal{E}}  - E_\mathrm{SFAM}^{\mathcal{Q}}
        \label{eq:sfam:me}
\end{equation}
The latter part subtracts all contributions that are covered by the QM calculation to avoid double counting. For EE, the QM/SFAM energy expression reads
\begin{equation}
        E_\mathrm{QM/SFAM}^\mathrm{EE} = E_\mathrm{QM/SFAM}^\mathrm{ME} + E_{ \mathrm{QM, estat}}^\mathcal{Q} - \sum_{\substack{A \in \mathcal{Q} \\ B \in \mathcal{E}}} E_\mathrm{estat}^{AB},
\end{equation}
with the QM electronic energy obtained with electrostatic embedding.
To account for the unfavorable scaling of the error in the total energy for large molecular systems, we proposed a reduced QM/MM energy:
\begin{equation}
        E_\mathrm{QM/SFAM}^\mathrm{red} = E_{QM}^\mathcal{Q} - E^{\mathcal{Q}-\mathcal{E}},
        \label{eq:reduced_qmmm}
\end{equation}
where $ E^{\mathcal{Q}-\mathcal{E}}$ is the interaction energy of either ME or EE type that is captured by the QM method.
This formalism neglects covalent SFAM contributions and noncovalent interactions between environment atoms, and therefore, it compensates the potentially large uncertainties induced by large environments $\mathcal{E}$.

QM/SFAM also provides a universal procedure to select an accurate QM region as the reactive center. To this end, candidate QM regions around a central atom are generated leveraging the SFAM fragmentation approach coupled to a stochastic element (with bonds being cut with some probability $p$). The candidate regions are selected within an upper and lower boundary in terms of atom count. These parameters allow to flexibly define the QM region size and variation thereof.
For that purpose, the mean absolute error of the force components on atoms that are closer to the selected central atom than a give threshold $r_\text{repr}$ in the candidate regions with respect to the larger reference regions are calculated.
The mean absolute error of one atom $k$ with regard to the reference is defined as
\begin{equation}
        e_k^m = \frac{1}{3} \Big( |f_{x, \text{ref}, k} - f_{x,m,k}| + |f_{y, \text{ref}, k} - f_{y,m,k}| + |f_{z, \text{ref}, k} - f_{z,m,k}|\Big),
        \label{eq:mean_on_one_atom}
\end{equation}
and the mean of all representative atoms as
\begin{equation}
        e_\text{mean}^k = \frac{1}{N_\text{repr}}\sum_{k=1}^{N_\text{repr}} e_k^m.
        \label{eq:symmetry_measure}
\end{equation}
In Eq.~(\ref{eq:mean_on_one_atom}), $\mathbf{f}_{\text{ref}, k} = (f_{x, \text{ref}, k} , f_{y, \text{ref}, k} , f_{z, \text{ref}, k} )^T$ is the reference force vector of atom $k$ and can be calculated with any QM method. The generated candidate regions can also be scored by the number of cleaved bonds at the QM/MM boundary (and hence, the number of inserted link atoms), and by the overall symmetry of the QM region.

To employ QM/SFAM for optimization routines provided within the SCINE infrastructure, we provide dedicated QM/MM optimizers that find local minima and QM/MM transition states on a highly complex potential energy surface. Both optimizers rely on a microiterative structure: For structure optimizations, all Cartesian coordinates of the quantum region and those of the environment that are bound to the quantum region are constrained, while all remaining MM degrees of freedom are relaxed until convergence (or until a maximum number of iterations is reached). After that, the complete system is optimized to a local minimum applying the full analytical QM/SFAM gradients.

For transition state optimizations, we exploit a partial Hessian scheme to optimize all atoms in the QM region while, at the same time, converging the atoms in the MM region to a local minimum. This approach avoids the calculation of MM Hessians. All MM atom positions are iteratively updated  as point charges in the QM Hamiltonian after convergence of the MM optimization. Then, the QM region is optimized by applying a Bofill update (eigenvector following), and the QM positions are updated in the full model after the maximum number of iterations. As the subsystems are optimized with individual optimizers, convergence is achieved when the QM optimization converges twice (before and after an MM optimization) in order to guarantee self-consistency.

\subsubsection{I/O}
The algorithms for structure preparation, model generation, and quantum region selection can be carried out with Swoose with an input file following the YAML syntax, which allows one to specify all required settings in a straightforward way. Additionally, Swoose can carry out single-point energy calculations, structure optimizations, and molecular dynamics simulations directly on the command line.

\subsubsection{API}
The Python bindings of the Swoose module provide the same functionality as the executable.
The Python module consists of functions that take a single structure file and optional keyword arguments.
Additionally, it includes the \texttt{Parametrizer} and \texttt{QmRegionSelector} classes (which can carry out the parametrization and quantum region selection tasks), gives access to the task-specific settings and to additional data structures related to these tasks.

\subsection{Parrot}

\subsubsection{Field of application}
Machine learning potentials offer a way to combine accuracy and efficiency in energy and gradient calculations. Their training is typically based on accurate data from electronic structure calculations. Machine learning potentials can preserve this accuracy, while their evaluation inflicts little computational demands comparable to those of force fields. In this way, machine learning potentials can enable accurate high-throughput screening. Parrot provides an interface to machine learning potential predictions of energy and gradients for a chemical structure defined by its elements and atomic positions. In addition, Parrot enables numerical Hessian calculations. Atomic charges and bond orders are obtained by GFN2-xTB\cite{Bannwarth2019} if the machine learning method is not able to predict these data.

\subsubsection{I/O}
Parrot interfaces APIs of established pretrained machine learning potential methods (currently TorchANI\cite{Gao2020}, Materials Graph Library\cite{mgl053}, and MACE\cite{Batatia2022}). Reading of model parameter files is handled by the APIs of the respective machine learning potential. Moreover, Parrot allows for reading and predictions of lifelong machine learning potentials\cite{Eckhoff2023}. Training and saving of improved lifelong machine learning potentials is not yet available in Parrot.

\subsubsection{API}
Parrot enables the selection of machine learning potentials as potential-energy models in SCINE in the same way as it is the case for other methods such as DFT, semi-empirical methods, and so forth. An overview of all machine learning potentials available in Parrot is provided in Table~\ref{t:parrot-methods}.

\begin{table}[htb!]
\caption{Machine learning potentials available in Parrot.}
\begin{center}
\begin{tabular}{ll}
\hline\hline
Method Family & Method \\
\hline
ani \cite{Gao2020} & ani2x \cite{Devereux2020} \\
ani \cite{Gao2020} & ani1ccx \cite{Smith2019} \\
ani \cite{Gao2020} & ani1x \cite{Smith2018} \\
m3gnet \cite{mgl053} & m3gnet-mp-2021.2.8-pes \cite{Chen2022} \\
m3gnet \cite{mgl053} & m3gnet-mp-2021.2.8-direct-pes \cite{Chen2022} \\
mace \cite{Batatia2022} & mace-mp{\_}large \cite{Batatia2023} \\
mace \cite{Batatia2022} & mace-mp{\_}medium \cite{Batatia2023} \\
mace \cite{Batatia2022} & mace-mp{\_}small \cite{Batatia2023} \\
mace \cite{Batatia2022} & mace-off{\_}large \cite{Kovacs2023} \\
mace \cite{Batatia2022} & mace-off{\_}medium \cite{Kovacs2023} \\
mace \cite{Batatia2022} & mace-off{\_}small \cite{Kovacs2023} \\
lmlp \cite{Eckhoff2023} & <user-defined> \cite{Eckhoff2023} \\
\hline\hline
\end{tabular}
\end{center}
\label{t:parrot-methods}
\end{table}

\subsection{AutoCAS}

\subsubsection{Field of application}
Highly accurate electronic structure calculations require one to take electron correlation into consideration. Even though closed-shell molecules can be routinely and reliably calculated with black-box coupled-cluster models\cite{Bartlett2024}, strongly correlated systems require an active orbital space for a multi-configurational description with subsequent treatment of the dynamic correlation. Since the manual selection of an active space requires chemical intuition and experience, this step is often time consuming, difficult to reproduce, and hardly possible for non-prototypical complex system such as the one in Ref.~\citenum{Sinha2017}. 
Especially for methods which can handle large active spaces of more than 100 orbitals, like DMRG\cite{White92, White93, 
Baiardi2020} and full configuration interaction quantum Monte Carlo\cite{Alavi2009,Cleland2010}, a manual selection is neither feasible nor reliable. Although many approaches\cite{Pulay1988, Bofill1989, Tishchenko2008, Bao2016, Bao2017, Sayfutyarova2017, Bao2018, Khedkar2019, Sayfutyarova2019, Jeong2020, Li2020, Gieseking2021, Golub2021, Khedkar2021, Lei2021, Levine2021, Oakley2021, King2021, Weser2022, Casanova2022, Cheng2022, King2022, Kaufold2023, Golub2023},
 have been devised to address this problem (at least in parts), we have demonstrated in 2016 a viable, reliable, and truly fully automated selection approach based on quantum information measures, the autoCAS approach\cite{Stein2016, Stein2016a, Stein2017a, Stein2019, Bensberg2023b}, 
 a development, which has not been paralleled so far.

The autoCAS algorithm\cite{Stein2016a, Stein2019} exploits the fact that DMRG can converge a qualitatively correct wave function with less effort than what would be required for a converged energy\cite{Moritz2007}.
Electron correlation encoded in this (approximate) wave function can be efficiently dissected in terms of orbital entanglement measures\cite{Boguslawski2012}. Hence an approximate DMRG calculation for all valence orbitals (including also double shells such as $4d$ orbitals for $3d$ transition metal atoms) delivers approximate, but qualitatively correct one- and two-orbital reduced density matrices, from which the single orbital entropies and the mutual information\cite{Legeza2003, Rissler2006, Legeza2006} are evaluated. Based on the single-orbital entropies, autoCAS defines the orbitals for an active space in a following fully converged CASSCF or DMRG calculation, optionally with subsequent dynamic correlation treatment. Even though the autoCAS algorithm is computationally more expensive than other active space selection algorithms, it is based on a multi-configurational description of the wave function and it is therefore not biased toward a single configuration, by contrast to other approaches. Moreover, since the approximate initial DMRG calculation relies on a rather small bond dimension and only a few sweeps to be efficient, the final fully converged calculation allows one to assess whether the initial settings were reliable enough for the active space selection. Should a discrepancy be detected because of truly strong correlations in the electronic structure, then one can easily restart the whole procedure with an increased bond dimension.

\subsubsection{I/O}
The autoCAS program package\cite{autocas210} provides different workflows for black-box active space calculations. Every workflow starts with a Hartree--Fock calculation, a subsequent active space selection, and a final calculation with the automatically selected active space. For small or test calculations the command-line interface provides the options to specify a basis set and a XYZ file containing the molecular structure. However, since the command-line interface is limited, production runs should use a YAML configuration file.  Especially crucial options, such as the large active space protocol\cite{Stein2019} and the required number of orbitals in each sub-space, can be specified there. 

A user-friendly option to interact with autoCAS is provided by the graphical user interface Heron\cite{heron100}. Heron does not only provide more control over autoCAS, which is especially helpful for an unexperienced user, it also features a built-in orbital viewer. The true power of autoCAS, however, comes by including the active space selection in custom workflows through the API.

\subsubsection{API}
In the design of autoCAS, we focused on a strict separation of the algorithms and the electronic structure backend. An example autoCAS script could have the following form:
\begin{lstlisting}[language=Python, caption=Minimal autoCAS workflow with the OpenMolcas interface.]
from scine_autocas import Autocas
from scine_autocas.autocas_utils.molecule import Molecule
from scine_autocas.interfaces.molcas import Molcas

xyz_file = "/path/to/molecule.xyz"
# Create a molecule
mol = Molecule(xyz_file)
ac = Autocas(mol)
# Initialize interface
molcas = Molcas([mol])
molcas.settings.basis_set = "cc-pVDZ"
molcas.settings.xyz_file = xyz_file

# Basic Workflow
occ, ind = ac.make_initial_active_space()
# HF calculation
molcas.calculate()
# CAS calculation
e, s1, s2, mi = molcas.calculate(occ, ind)
occ, ind = ac.get_active_space(occ, s1)
# Final calculation
e, s1, s2, mi = molcas.calculate(occ, ind)
\end{lstlisting}

The \texttt{Molecule} class stores all molecule-specific values, such as spin and charge. It can be initialized through a XYZ file or a list of atoms (strings with element labels). The \texttt{Autocas} class is responsible for the handling of active spaces, either the valence space (number of occupied and virtual orbitals around the Fermi vacuum) or the selected active space based on single-orbital entropies. These functions return 2 lists: one list with the orbital indices and one list storing an occupation number for each orbital with a possible value of 2, 1 or 0, representing a doubly occupied, singly occupied or virtual orbital, respectively. Hence, the total number of electrons in an active space can be determined by summation of the entries in the occupation number list.  
Even though every interface in autoCAS inherits from a base class, customized workflows only need to provide a list with single-orbital entropies in the same order as the index and occupation lists.

\subsection{ReaDuct}

\subsubsection{Field of application}
\label{sssection:readuct_field_of_application}

Chemical reaction exploration relies on the determination of minima and first-order saddle points on potential energy surfaces. Minima correspond to stable molecular structures, while first-order saddle points represent transition state (TS) structures, which connect minima in reaction valleys. The tasks required to locate and connect these points on a PES can be performed by the SCINE module ReaDuct\cite{Vaucher2018, readuct510}. ReaDuct is designed in such a way that the execution of a task is entirely independent of the method characterizing the PES and providing the necessary quantum chemical properties, such as energies and nuclear gradients.

A typical workflow for discovering new reactions involves, starting from one (single-ended) or two (double-ended) minima, an initial transition state search task followed by a transition state optimization task. The optimized TS structure can then be associated to an elementary step by an intrinsic reaction coordinate (IRC) task, whose resulting endpoints can be optimized with a structure optimization.

Each task offers various options and is highly adaptable to the chemical problem at hand. A comprehensive and detailed description of all options can be found in the manual of the module. A few key options are listed in the following. For instance, given one minimum structure, TS guess structures to another minimum can be found with a transition state search task employing our Newton Trajectory Algorithm 1\cite{Unsleber2022}, Algorithm 2\cite{Unsleber2022}, and AFIR algorithm developed by the Maeda group\cite{Maeda2010, Maeda2011, Maeda2014}. For the TS structure optimization, Hessian-based algorithms such as Eigenvector following\cite{Banerjee1985, Schlegel2011}, an approach proposed by Bofill\cite{Bofill1994} or a solely gradient-based algorithm\cite{Henkelman1999, Kaestner2008, Shang2010} are available. To connect a TS structure to reaction valleys, an IRC calculation along the eigenvector of the lowest eigenvalue of the Hessian, employing a steepest decent optimizer, can be executed. The resulting endpoints can be further optimized with various optimizers, which either explicitly calculate or approximate the Hessian, to obtain the local minima on the PES connected by the previously determined TS structure.

\subsubsection{Scientific background}

Some of the algorithms accessible through ReaDuct are well-established in the literature, while others were specifically designed within the SCINE project.
One example of the latter are the so-called Newton trajectory tasks, originally introduced in Ref.~\citenum{Unsleber2022} and later improved upon in Ref.~\citenum{Steiner2024}.
Their development was inspired by the work of Quapp and Bofill on Newton trajectories\cite{Quapp2020} (hence the name) and their analysis of Maeda's artificial force induced reaction (AFIR) approach\cite{Maeda2010, Maeda2011, Maeda2014}. The Newton Trajectory task~2~(NT2) algorithm has become our workhorse for single-ended transition state searches, which is why we describe the underlying algorithm in the following.

\begin{figure}[H]
    \centering
    \includegraphics[width=0.82\columnwidth]{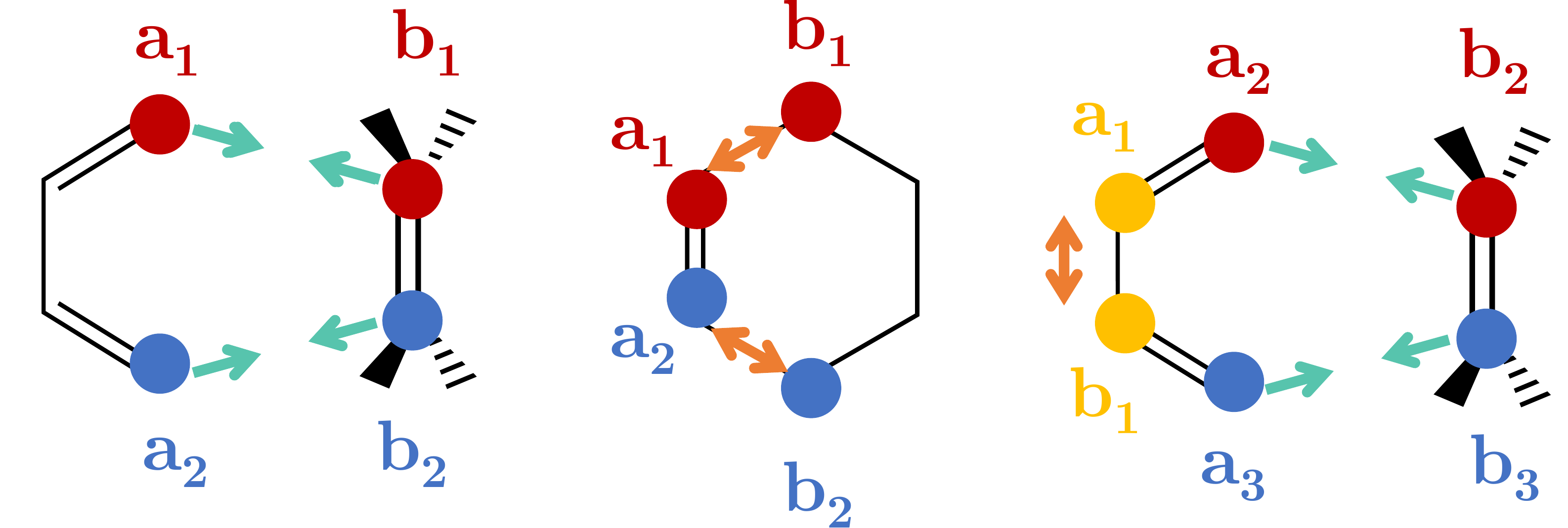}
    \caption{Reactive coordinates in NT2 tasks. Reactive pairs (two circles of same color) can be associative (green arrows) or dissociative (orange arrows). Reactive coordinates may include associative or dissociative reactive pairs only (left and center) or both types of pairs simultaneously (right). Figure adapted from Ref.~\citenum{Grimmel2022}.}
    \label{fig:nt2}
\end{figure}

Starting from a reactant's minimum energy structure (or a reactive complex composed of several reactants) NT2 scans aim to locate transition state guesses in a given direction. This direction is specified by a list of so-called reactive atom pairs, referred to as the reactive coordinate. Each pair is either set to be pushed together (associative) or to be pulled apart (dissociative). For an illustration see Fig.~\ref{fig:nt2}.
Note that while the selection of these pairs can be motivated by the expectation of certain bonds to be formed or broken, the algorithm neither strictly enforces specific bond modifications nor does it limit changes in the molecular graph to the selected atom pairs.

An NT2 scan is a steepest descent~(SD) structure optimization under the application of a modified gradient, such that the system is pushed energetically uphill into the direction of interest.
This modified gradient $\mathbf{g}^{\text{NT2}}$ is obtained from the original electronic energy gradient $\mathbf{g}$ as follows: While the gradient components acting on atoms that are not included in the reactive coordinate remain unchanged,  the gradient of each reactive atom $k$ is orthogonalized with respect to the vectors $\mathbf{R}_{a_i b_i}$ connecting the reactive atom pairs $\mathbf{P}^{\text{react}}_i = (a_i, b_i)$ to which $k$ belongs:
\begin{equation}
    \mathbf{g}_k \rightarrow \textbf{g}_k^\perp,\quad \text{with}\quad\textbf{g}_k^\perp\perp \textbf{R}_{a_i b_i} \quad \forall \textbf{P}_i^{\text{react}} = (a_i, b_i) \ni k
\end{equation}
Subsequently, artificial force components are included,
\begin{equation}
    \mathbf{g}_k^\text{NT2} =  \textbf{g}_k^\perp - 0.5 \frac{\alpha_\text{NT2}}{\text{max}_{\forall (a_i, b_i) \ni k} || \mathbf{X}_{a_i b_i} ||} \cdot \left( \sum_i \mathbf{X}_{k b_i} - \sum_i \mathbf{X}_{a_i k} \right),
\end{equation}
where $\alpha_\text{NT2}$ is a positive force constant.
The two sums run over all reactive pairs with atom $k$ being their first or second element.
The calculation of $\mathbf{X}_{a_i b_i}$ depends on whether the pair $(a_i, b_i)$ was set to be pushed together or to be pulled apart.
For the associative case $\mathbf{X}_{a_i b_i}$ is
\begin{equation}
    \mathbf{X}_{a_i b_i}^\text{associative} = \left( || \mathbf{R}_{a_i b_i}|| - \left( r_{a_i}^{\text{cov}} + r_{b_i}^{\text{cov}}\right)\right) \cdot \frac{\mathbf{R}_{a_ib_i}}{||\mathbf{R}_{a_ib_i}||}
\end{equation}
with $||\mathbf{R}_{a_ib_i}||$ denoting the distance between atoms $a_i$ and $b_i$ and $r_{a_i}^{\text{cov}}$ and $r_{b_i}^{\text{cov}}$ their covalent radii, respectively.
Hence, the prefactor $|| \mathbf{R}_{a_i b_i}|| - \left( r_{a_i}^{\text{cov}} + r_{b_i}^{\text{cov}}\right)$ scales the force component according to the difference between the actual interatomic distance and an idealized bond distance.
This facilitates concerted bond formations when multiple associative pairs are considered simultaneously.
For pairs that are to be pulled apart $\mathbf{X}_{a_i b_i}$ is the reversed, normalized intra-pair connection vector scaled with a factor $\lambda$,
\begin{equation}
    \mathbf{X}_{a_i b_i}^\text{dissociative} = - \lambda \cdot \frac{\mathbf{R}_{a_ib_i}}{||\mathbf{R}_{a_ib_i}||}
\end{equation}
Adjusting $\lambda$ allows one to control the magnitude of associative vs. dissociative components of the overall artificial force.

Each SD step is followed by BFGS optimization cycles during which the positions of all atoms that are part of the reactive coordinate are frozen, while the remainder of the structure relaxes, \textit{i.e.}, the remainder adapts to the induced changes.
One SD step and the subsequent BFGS relaxation steps are referred to as a macrocycle.

The scan is stopped (i) upon reaching a user-specified number of gradient calculations or (ii) if the distance between reactive pairs being pushed together is smaller than the sum of their covalent radii times 0.9 or the bond order between them exceeds 0.75, and the bond order between atom pairs that are pulled apart is below 0.15. Furthermore, the scan finishes when a gradient calculation fails.
A transition state guess structure is extracted from the trajectory obtained from the macrocycles of the scan (\textit{i.e.}, post-BFGS relaxation)
based on changes in the bonding pattern within the reactive pairs and the electronic energy curve (for details see Ref.~\citenum{Steiner2024}).
The obtained guess structure can then be subjected to a transition state optimization as outlined in section~\ref{sssection:readuct_field_of_application}.

\subsubsection{I/O}

All functionality of ReaDuct can be accessed via an input file following the YAML syntax. The input file comprises two blocks: one defining multiple systems with their nuclear coordinates and the method defining the PES. The second block comprises the requested tasks with the relevant settings. Multiple tasks can be added and are executed after each other. The ReaDuct program is then simply run from the command line with
\begin{lstlisting}[language=Bash, label=code:readuct_run]
readuct -c input.yaml
\end{lstlisting}

\subsubsection{API}

The ReaDuct module seamlessly integrates into workflows through its Python bindings. Executing tasks differs slightly from directly calling the ReaDuct binary. Instead of a single block for the system in an input file, one utilizes the Python bindings of our SCINE Utilities to create the corresponding Python object for the system (calculator).

Various tasks can then be executed through simple function calls, with the system (calculator) as an argument. The function returns a boolean (indicating whether the task was successful or not) and a Python dictionary containing the calculator with its results:

\begin{lstlisting}[language=Python, label=code:python_readuct_bo, caption=Performing a bond order calculation task with ReaDuct.]
import scine_utilities as utils
import scine_readuct as readuct
import scine_sparrow

water_calculator = utils.core.load_system("h2o.xyz", "PM6")

systems = {}
systems["water"] = water_calculator
systems, success = readuct.run_bond_order_task(systems, ["water"])
results = systems["water"].get_results()
print("Energy", results.energy)
\end{lstlisting}

The \texttt{results} Python object may have different attributes, such as the energy or the bond orders. Task-specific settings are provided as keyword arguments for the ReaDuct functions.

\begin{lstlisting}[language=Python, label=code:python_readuct_go, caption=Performing a structure optimization task with ReaDuct.]
optimization_settings = {"output": ["water_opt"], "optimizer": "BFGS"}
systems, success = readuct.run_optimization_task(systems, ["water"], **optimization_settings)
opt_results = systems["water_opt"].get_results()
print("Opt. Energy", results.energy)
\end{lstlisting}

This allows for sequential execution of various tasks, with results analyzed in a straightforward Python workflow. A prime example of such workflows is demonstrated in the numerous jobs within our SCINE Puffin module (see next section).

\subsection{Puffin}
\label{sec:puffin}

\subsubsection{Field of application}
\begin{figure}[H]
    \centering
    \includegraphics[width=0.82\columnwidth]{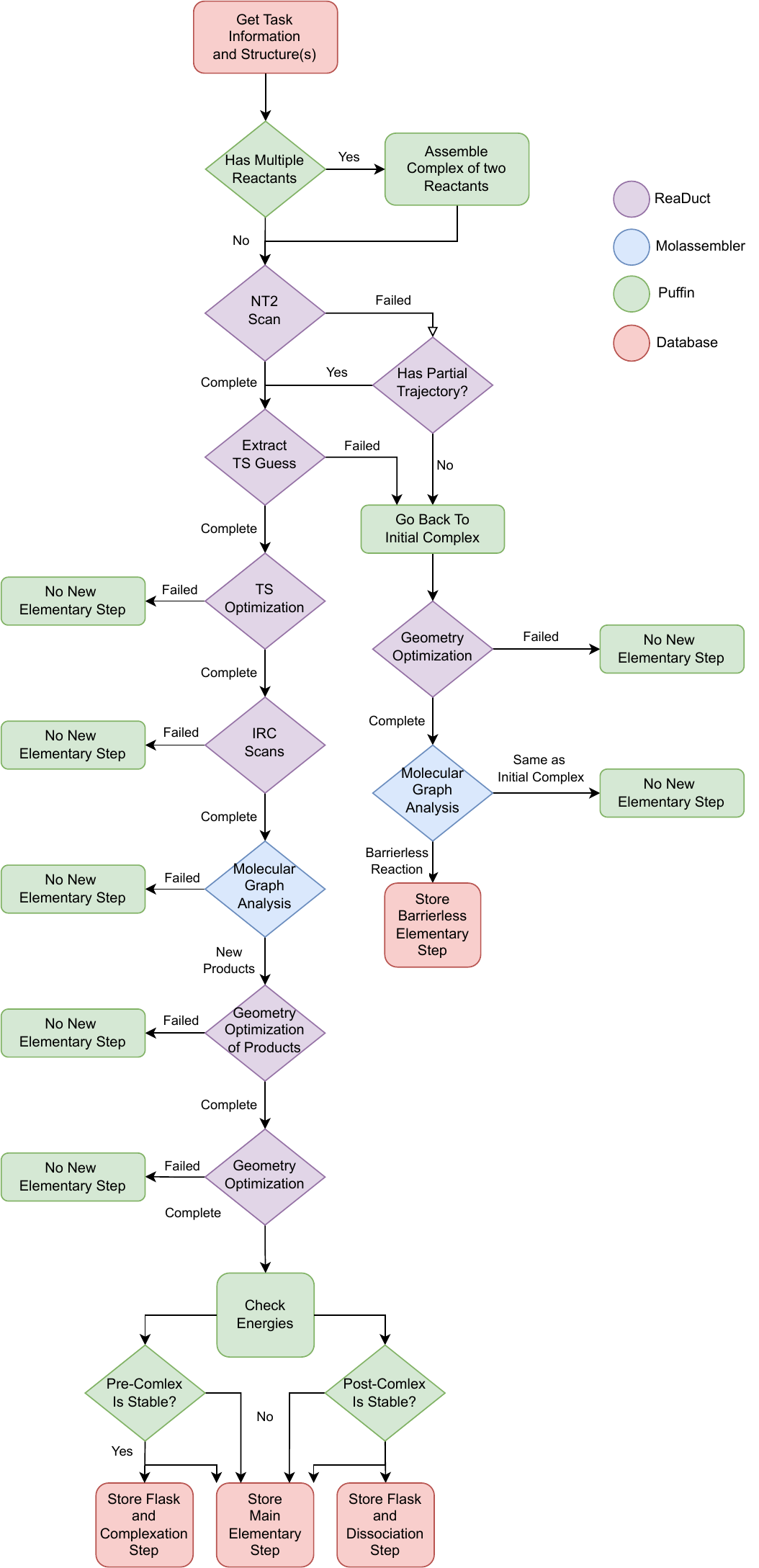}
    \caption{Flowchart of an elementary-step search as implemented in SCINE Puffin.}
    \label{fig:reaction_trial_schema}
\end{figure}

Puffin's main aim is to combine basic algorithms implemented in dedicated electronic structure packages or abstracted by other SCINE modules into larger tasks representing a single calculation in network exploration. These calculations can ultimately be simple tasks, such as calculating a single-point energy for a given structure or, as seen in Figure~\ref{fig:reaction_trial_schema}, they can also be rather involved.
Figure~\ref{fig:reaction_trial_schema} shows the diagram of a single elementary-step trial, including fall-back strategies and alternative branches based on intermediate results. Multiple programs and a direct connection to the exploration database are used. Those steps marked to be carried out by ReadDuct can use any electronic structure program interfaced as a SCINE Core Calculator.

Puffin is a program that is launched in high numbers on high-performance computing infrastructure in order carry out composite workflows. At startup, the user can assign the computational resources (such as the number of processors and the amount of memory) which a Puffin instance is allowed to utilize. These resource limits will subsequently be propagated by Puffin to any further program it invokes. SCINE Puffin is currently employed in, but not limited to, our exploration framework as it can encode any algorithm that requires no more than a set of atomistic input structures and an electronic structure model. Therefore, possible fields of application are high-throughput campaigns that carry out individual energy evaluations, optimization algorithms, or more complex workflows based on chained optimization algorithms such as reaction trials that combine reactive complex formation, Newton trajectory, single-ended transition state search, intrinsic reaction coordinate calculations, and additional checks to assess the bonding patterns, molecular charge and spin multiplicity of any found product(s).

To carry out these workflows, Puffin depends on other SCINE modules such as ReaDuct and possibly external quantum chemistry software. To simplify the setup of this large software stack, we built a so-called bootstrapping procedure through which a Puffin instance automatically downloads and compiles all required software. With this bootstrap procedure, installing the entire SCINE toolchain becomes straightforward even for unexperienced users. Large-scale deployments on high-throughput computing infrastructure are further simplified by running Puffin in a virtualized environment. For this, we currently support Docker and Apptainer images.

\subsubsection{API}
A new workflow can be defined by defining a new \texttt{Job}. A job is generally defined by its unique name that signals to a Puffin instance which workflow it must execute, and a \texttt{run} method that is carried out during execution in between general jobs, resource, and data management.

On top of that, Puffin automatically manages the different available resources, programs, and program versions so that it only carries out workflows appropriate to a specific instance and logs the employed programs. One computational resource that is difficult to estimate, especially in high-throughput settings, is the memory usage. Therefore, each instance receives a maximum memory requirement. During the workflow execution, Puffin monitors itself to stay below this limit and aborts the operation otherwise.

\subsection{Database}
\label{sec:db_wrapper}

\subsubsection{Field of application}
The database plays a central role in our automated explorations as it is not only the permanent data storage, but also the mean of communication between the software driving the exploration forward, SCINE Chemoton\cite{Unsleber2022, chemoton310}, and the software executing individual calculations, SCINE Puffin\cite{puffin130}. 

The sheer size of chemical reaction networks dictates to store the exploration progress in a database that can be scaled to millions of entries. Furthermore, the quantum chemical exploration of CRNs generates additional information about the electronic structure of individual chemical structures and minimum energy paths, which can be stored for subsequent campaigns to build data-driven models\cite{Simm2018, Gugler2022, Eckhoff2023}.

In case the exploration is carried out autonomously, \textit{i.e.}, calculations are set up iteratively based on the already explored reaction space, it is desirable to employ the same database for both, carrying out calculations and storing the CRN. However, this requires the database to be able to handle thousands of simultaneous queries from the algorithms driving the exploration and the processes that carry out individual electronic structure calculations. Hence, we have developed a database schema for a MongoDB\cite{MongoDB2023} database and a custom wrapper to interact with the database\cite{database130}. 
This wrapper provides useful functionality geared towards the exploration of CRNs which is not available from MongoDB itself. For example, entities such as reactions and structures are defined by the wrapper, as well as related functions (such as retrieving the atoms constituting a given structure). The wrapper has been designed such that every interaction with the database is synchronized with its current state. This guarantees that multiple processes can act on the current database simultaneously.

MongoDB databases consist of separate documents. In our database schema, different types of documents are defined, which are stored in collections of identical document types. The database encodes the CRN by linking several documents across multiple collections. Additionally, information about the carried out calculations and their results are stored in separate collections. The different document types are defined as:
\begin{itemize}
\item \textbf{Structure:} a chemical structure defined by Cartesian coordinates of atoms, molecular charge, and spin multiplicity, representing a point on a Born--Oppenheimer potential energy surface
\item \textbf{Property:} a quantum chemical property of a specific structure
\item \textbf{Elementary Step:} a transformation of a set of structures to a different set of structures with an allowed intersection of the two sets
\item \textbf{Calculation:} a sequence of individual quantum chemical calculations on a set of structures that can generate new structures, properties, and elementary steps
\item \textbf{Compound:} a collection of structures that have an identical element composition, molecular charge, spin multiplicity, connectivity, and local arrangement around each nucleus. The latter is determined by the shape fitting and ranking algorithms in SCINE Molassembler\cite{Sobez2020, molassembler201}
\item \textbf{Flask:} identical to Compound, but the contained structures are not required to be connected in one single graph
\item \textbf{Reaction:} a collection of elementary steps that transform structures belonging to identical compounds or flasks
\end{itemize}

\begin{figure}[H]
    \centering
    \includegraphics[width=0.82\columnwidth]{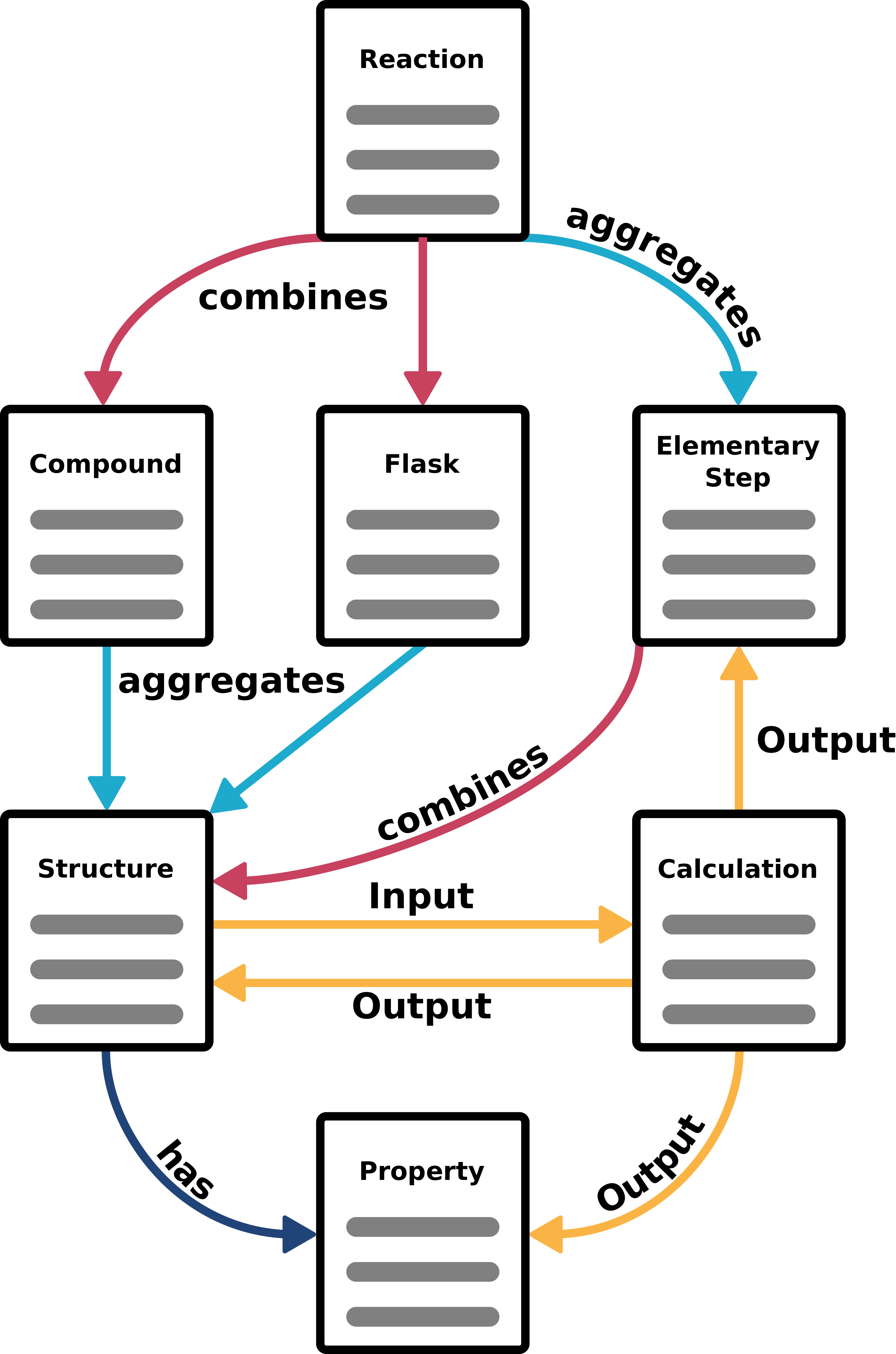}
    \caption{Schema of the SCINE Database in a MongoDB format. The document symbols show the different types of documents, which are stored in one collection each. The relationships between the documents are facilitated through cross references by unique identifiers in each document.}
    \label{fig:database_schema}
\end{figure}

We refer to Ref.~\citenum{Unsleber2020} for more details on the individual definitions. The relations between all documents in the SCINE Database are illustrated in Fig.~\ref{fig:database_schema}.

\subsubsection{API}
The SCINE Database is built around the general MongoDB database. In order to run a SCINE Database a MongoDB instance must be started and be addressable with an IP address, port, database name, and optional authentication credentials. Since SCINE Database is a C++ library, the database can also be addressed in C++ programs. However, our library also features Python bindings with which we showcase its capabilities here. We connect to the database with the \texttt{Manager} class:
\begin{lstlisting}[language=Python, label=code:database_connection, caption=Connecting to the SCINE database and accessing different collections.]
import scine_database as db

manager = db.Manager()
manager.set_credentials(db.Credentials("localhost", 27017, "database_name"))
# The following line will fail if you don't have
# a database running or incorrect credentials
manager.connect()
# If you are creating a database from scratch
# you can run
#    manager.init()
# to create the collections accessed in the
# following
calculations = manager.get_collection("calculations")
compounds = manager.get_collection("compounds")
flasks = manager.get_collection("flasks")
elementary_steps = manager.get_collection("elementary_steps")
reactions = manager.get_collection("reactions")
properties = manager.get_collection("properties")
structures = manager.get_collection("structures")
\end{lstlisting}

In order to retrieve information from the database, the collections objects hold the querying methods  \texttt{query\_x}, \texttt{iterate\_x}, and \texttt{count}, \texttt{x} being a placeholder for the applied collection such as \texttt{query\_calculations} or \texttt{query\_structures}. The querying methods (\textit{cf.}, Listing~\ref{code:database_simple_queries}) can take any valid MongoDB query, which are given as documents that specify which keys of the database documents should fulfill certain criteria, \textit{e.g.}, equality of values. More complex matches are achieved with operators, which start with \$, and multiple queries can be combined with logical operators such as ``\texttt{and}''/``\texttt{or}''.

\begin{lstlisting}[language=Python, label=code:database_simple_queries, caption=Overview of the basic querying method with the SCINE database.]
from json import dumps

# Count everything in the calculations
# collection
print("Total nr. of calculations:", calculations.count("{}"))
# Count failed calculations
print("Nr. of failed calculations:", calculations.count(dumps({"status": "failed"})))
# Get calculation IDs of all successful DFT
# calculations
selection = {"$and": [
    {"status": {"$in": ["complete", "analyzed"]}},
    {"model.method_family": "DFT"}
  ]
}
\end{lstlisting}

The underlying electronic structure methods can generally be summarized in the \texttt{Model} objects, which can be directly queried for. The results of the queries can be built directly into loops to aggregate information. For example, one can collect the Cartesian coordinates of all structures that were optimized with GFN2-xTB, but have already received a single-point energy obtained with an exchange--correlation density functional:
\begin{lstlisting}[language=Python, label=code:database_loop_query, caption=Example for querying all structures of a certain model and aggregating their coordinates based on additional criteria and their electronic energy with a different model.]
from scine_database.queries import model_query, optimized_labels

structure_model = db.Model("GFN2", "GFN2", "")
structure_model.solvent = "water"
structure_model.solvation = "any"
energy_model = db.Model("DFT", "PBE-D3BJ", "def2-SVP")

results = {}
for structure in structures.query_structures(dumps(
    {"$and": [
        {"exploration_disabled": False},
        {"label": {"$in": optimized_labels()}}
    ] + model_query(structure_model)
    }
)):
    prop_ids = structure.query_properties({
        "electronic_energy",
        energy_model,
        properties
    })    
    if prop_ids:
        atoms = structure.get_atoms()
        # Take most recent energy
        energy_prop = db.NumberProperty(prop_ids[-1], properties)
        results[str(structure.id())] = {
            "elements": atoms.elements,
            "coords": atoms.positions,
            "energy": energy_prop.get_data()
        }
\end{lstlisting}

More detailed queries, such as the spread of the forward barriers of all elementary steps within a reaction, might require aggregated information that is not directly stored in the database, but can be calculated on the fly. In addition to the Python bindings, the SCINE Database library also features Python functions that carry out such aggregated queries which are required in reaction network explorations similar to the one in Listing~\ref{code:database_loop_query}.

\subsection{Chemoton}

\subsubsection{Field of Application}
The exhaustive elucidation of reaction mechanisms demands mapping all competitive elementary steps into a chemical reaction network, for which various automated approaches have been devised\cite{Sameera2016, Dewyer2017, Simm2019, Unsleber2020, Steiner2022, Baiardi2022, Ismail2022, Wen2023, Margraf2023}. Our SCINE Chemoton module facilitates the exploration of CRNs in a fully automated fashion based on first principles\cite{Bergeler2015, Simm2018, Unsleber2022}. When provided with at least one starting molecule, Chemoton systematically discovers all possible reactions originating from this molecule and its successors. This process thereby reveals the entire reaction network. The modularity of Chemoton enables this comprehensive exploration\cite{Unsleber2022}.

\subsubsection{Scientific background}

An automated exploration requires several distinct tasks, each carried out independently and in parallel. The execution must be identical for all tasks, with tasks requiring the same architecture but addressing different facets of the automated exploration. Within our framework, we designate the execution component as the \texttt{Engine} and the task-specific aspect as the \texttt{Gear}. Each \texttt{Engine} is paired with a \texttt{Gear} and can be run individually.  

\begin{figure}[H]
\centering
\includegraphics[width=0.95\columnwidth]{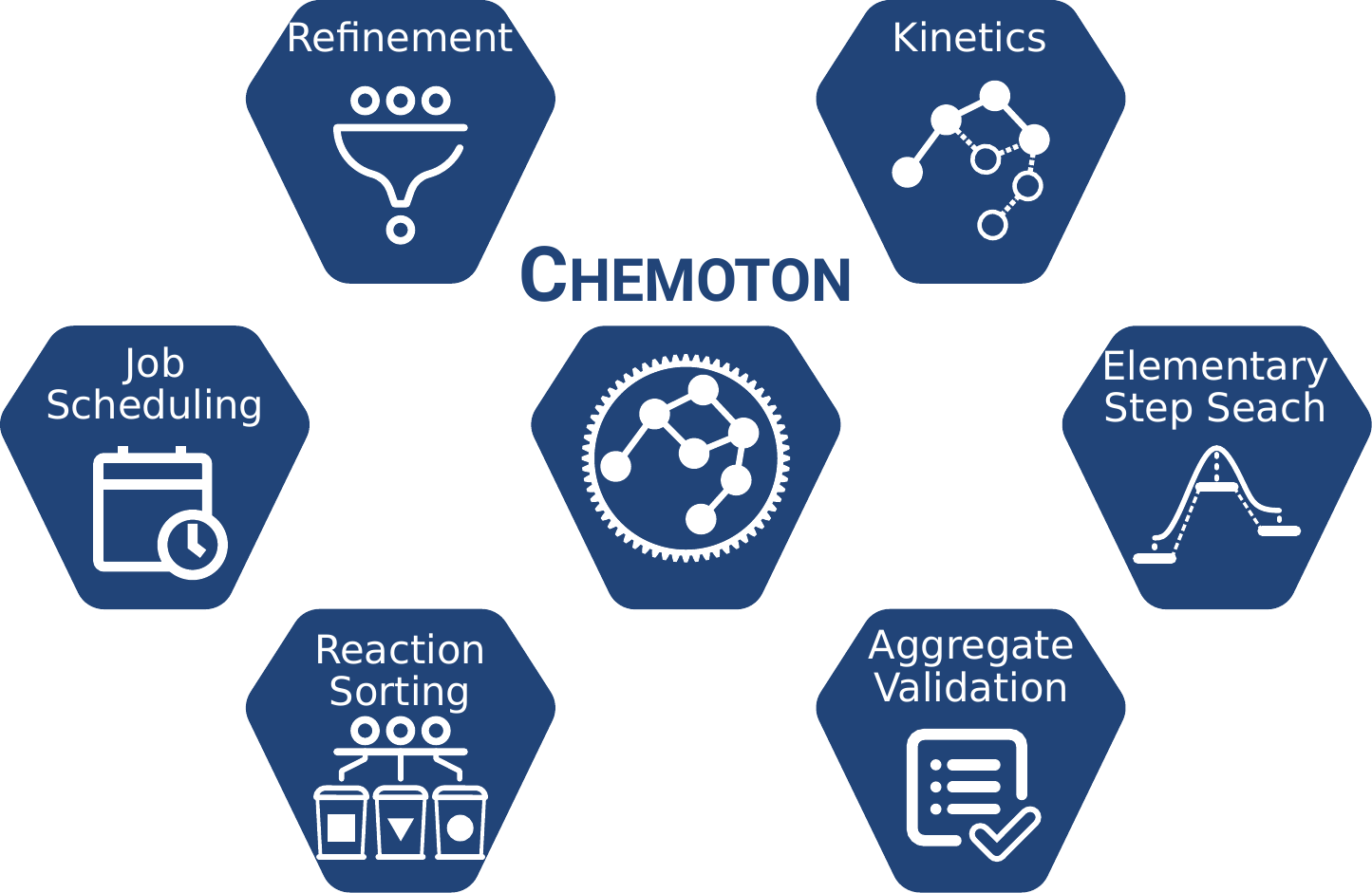}
\caption{Overview of Chemoton gears required for a chemical reaction network exploration.}
\label{fig:chemoton-overview}
\end{figure}

Key gears of the Chemoton workflow are depicted in Fig.~\ref{fig:chemoton-overview} and will be briefly explained in the following. The top-right gear in Fig.~\ref{fig:chemoton-overview}, the Kinetics gear, determines the molecules (aggregates) allowed for a reaction search. Moving clockwise, the Elementary Step Search gear probes one or two chemical species for a reaction, possibly in various orientations and along various reaction coordinates. The Aggregate Validation gear ensures that newly found molecules are local minima on their PES, sorting them into collections called ``aggregates'' of molecules based on multiplicity, charge, and bonding pattern. The latter is analyzed with Molassembler (see Section~\ref{sec:molassembler} for details). Similarly, the Reaction Sorting gear collects reactions connecting the same aggregates (see Section~\ref{sec:db_wrapper} for the nomenclature employed). To distribute the available resources based on each gear's needs, the Job Scheduling gear determines the order in which calculations should be executed. This guarantees a seamless exploration and full usage of resources. The Refinement gear allows for recalculating certain properties, such as the energy of a molecule, \textit{e.g.}, with a different electronic structure method.

The software structure presented so far is strongly tied to autonomous, indefinitely running, independently working processes, which is well suited for exhaustive explorations based on defined parameters for the complete exploration. However, if the chemical system or reaction mechanism is highly complex and does not lend itself easily to exhaustive explorations, a linear style of exploration can establish important nodes in the reaction network before expanding then on the paths obtained with automated approaches. Such a linear exploration is enabled by the top-level Steering Wheel framework\cite{Steiner2024} that combines several gears into discrete exploration steps. The exploration steps are either a \texttt{Network Expansion} that adds information to the reaction network or a \texttt{Selection Step} that defines a subset of the network. These must be applied in an alternating fashion which allows one to systematically steer an exploration into certain directions of chemical reaction space.

All calculations created by the running gears and required for an exploration are written to a database (Section~\ref{sec:db_wrapper}). Once they are in the database, they are simply jobs for our calculation backend Puffin (Section~\ref{sec:puffin}), responsible for running the jobs and writing their results into the database. For setting up new calculations and expanding the exploration, the running Chemoton gears are constantly checking on these results. Hence, all of the results and output created by running a chemical network exploration with Chemoton is stored in a database and chemical insights can be gained by querying this database.

The resulting networks are often too extensive and interconnected to be easily interpreted by a human being. Therefore, we developed Pathfinder within Chemoton to facilitate navigation and analysis of such networks\cite{Turtscher2023}. Pathfinder converts the network stored in the database (Section~\ref{sec:db_wrapper}) to a graph, encoding both stoichiometric and kinetic information in the edges. Given user-defined starting conditions, the aggregates in the graph have varying likelihood of being encountered based on their accessibility from a starting aggregate. The edges of the graph are then updated with this information, ensuring that unaccessible aggregates are disfavored in a simple shortest path search. In such a search, source and target vertices can be any vertices in the graph, providing a route in terms of reaction sequences for how a certain product is formed from a given starting aggregate under the specified starting conditions. Altering the starting conditions can lead to different shortest routes for a certain source--target vertex pair, allowing the investigation of various conditions. Overall, Pathfinder facilitates efficient navigation through and analysis of complex chemical reaction networks efficiently, providing intuitive insights for chemists in the form of reaction sequences and profiles.

\subsubsection{API}
Due to the scale of automated explorations, their setup depends strongly on the specific use case and desired properties. The most-approachable entry point is within our graphical user interface, which allows one to construct individual gears and offers pop-up menus for all their options. More experienced users can setup the individual gears in a small Python script as shown in Listing~\ref{code:chemoton_script}. The Chemoton main script serves as template for such scripts.

\begin{lstlisting}[language=Python, label=code:chemoton_script, caption=Example for running a Chemoton engine with a specific gear.]
from scine_chemoton.gears.thermo import BasicThermoDataCompletion
from scine_chemoton.gears.scheduler import Scheduler
from scine_chemoton.engine import Engine
from scine_database import Credentials

credentials = Credentials("localhost", 27017, "database_name")
# Hessian calculations
thermo_gear = BasicThermoDataCompletion()
thermo_engine = Engine(credentials, fork=False)
thermo_engine.set_gear(thermo_gear)
thermo_engine.run(single=True)
# Handle calculation scheduling
schedule_gear = Scheduler()
schedule_gear.options.job_counts["scine_hessian"] = 100_000
schedule_engine = Engine(credentials, fork=False)
schedule_engine.set_gear(schedule_gear)
schedule_engine.run(single=True)
\end{lstlisting}

A steered exploration can be carried out in a Python environment by constructing the linear exploration protocol. The example in Listing~\ref{code:steering} sets up an exploration that adds two compounds to the database, creates conformers, selects the lowest energy conformers in each cluster based on k-means clustering, and samples possible reactions between the compounds. Due to the linear execution logic of the Steering Wheel, it does not matter whether the two last steps were added later or from the beginning of the exploration run as both procedures produce identical results.

\begin{lstlisting}[language=Python, label=code:steering, caption=Carrying out a steered reaction exploration with Chemoton.]
from time import sleep
from scine_chemoton.steering_wheel import SteeringWheel
from scine_chemoton.steering_wheel.selections import AllCompoundsSelection
from scine_chemoton.steering_wheel.selections.input_selections import FileInputSelection
from scine_chemoton.steering_wheel.selections.conformers import LowestEnergyConformerPerClusterSelection
from scine_chemoton.steering_wheel.network_expansions.basics import SimpleOptimization
from scine_chemoton.steering_wheel.network_expansions.conformers import ConformerCreation
from scine_chemoton.steering_wheel.network_expansions.reactions import Association
from scine_chemoton.gears.elementary_steps.reaction_rules.reaction_rule_library import (
    DefaultOrganicChemistry,
    SimpleDistanceRule,
)
from scine_chemoton.gears.elementary_steps.reactive_site_filters import AtomRuleBasedFilter
import scine_database as db

model = db.Model("GFN2", "GFN2", "")
protocol = [
    FileInputSelection(
        model,
        [
        # Structure file, charge, spin mult.
        ["molecule_A.xyz", 0, 1],
        ["molecule_B.xyz", 0, 1]
    ]),
    SimpleOptimization(model, status_cycle_time=1),
    AllCompoundsSelection(model),
    ConformerCreation(model),
    LowestEnergyConformerPerClusterSelection(
        model,
        n_clusters=5,  # k-means
        additional_reactive_site_filters=AtomRuleBasedFilter(
            DefaultOrganicChemistry()
        )
    ),
    Association(
        model,
        max_bond_associations=2,
        max_bond_dissociations=1
    )
]
wheel = SteeringWheel(
    db.Credentials("localhost", 27017, "database_name"),
    protocol
)
wheel.run()  # runs non-blocking
sleep(1)
# One can query progress along the way:
print(wheel.get_status_report())
print("will run until all finished or stopped")
print("will only finish if calculations are executed")
try:
    while wheel.is_running():
        sleep(1)
except KeyboardInterrupt:
    wheel.terminate()
wheel.save()
\end{lstlisting}

Due to the top-level approach and flexibility of the Steering Wheel, the decision on the exploration protocol requires \textit{a priori} either a hypothesis about the possible reaction mechanism or a good overview of the status of the exploration in order to adapt the exploration strategy on the fly. Because the first requirement is rare, we have built the framework around the second requirement by integrating it into our graphical user interface Heron. This allows one to access the current exploration status and preview the effect of potential further exploration steps.

\subsection{KiNetX}

\subsubsection{Field of application}
Microkinetic modeling simulations are crucial to connect the microscopic description of a chemical reaction as a detailed reaction network to macroscopic observations such as yields by predicting the concentration fluxes through the network. Microkinetic modeling simulations can be run through the SCINE module KiNetX\cite{kinetx200}.
The version of KiNetX available in SCINE is written in modern C++ and is based on its original MATLAB implementation\cite{Proppe2019, Proppe2022}.
Microkinetic modeling simulations can automatically steer the reaction network exploration\cite{Bensberg2023} by interweaving them directly in a rolling exploration. KiNetX is interfaced with the library Sundials\cite{Hindmarsh2005, Gardner2022}, which provides efficient integration routines for the differential equations in the microkinetic modeling. As an alternative to KiNetX, SCINE provides an interface to the program Reaction Mechanism Simulator (RMS)\cite{Johnson2022a, Johnson2023} for microkinetic modeling. To predict which reactions and compounds in a reaction network dominate the overall kinetics and therefore must be described accurately, the SCINE framework provides sensitivity analysis approaches for the RMS microkinetic modeling through the library SALib\cite{Herman2017, Iwanaga2022}. The sensitivity analysis can also be interwoven with a rolling exploration to identify the key reactions and compounds and refine their descriptions by reoptimizing structures or recalculating electronic energies with more accurate electronic structure models\cite{Bensberg2023a}.

\subsubsection{I/O}
KiNetX assumes constant volume and temperature for the microkinetic modeling. It requires only rate constants, reactive species, reactions, start concentrations, and simulation time or number of time steps as input. During the simulation, it prints the concentrations for batches of time steps to the screen. The calculation input must be provided through its Python API.

\subsubsection{API}
KiNetX can be easily integrated into automated workflows through its Python bindings. First, the forward/backward rate constants (\texttt{k\_fs}/\texttt{k\_bs}), reaction stoichiometry (\texttt{s\_fs}/\texttt{s\_bs}), and the compounds must be encoded in a Python object that represents the reaction network. This is accomplished by successively adding compounds and reactions to a \texttt{NetworkBuilder} object:

\begin{lstlisting}[language=Python, caption={Encoding the reaction network for KiNetX (for the sake of brevity, this code listing cannot be executed as such, since variables such as \texttt{compound\_masses} are not defined).}]
import scine_kinetx as kx
network_builder = kx.NetworkBuilder()
for mass, name in zip(compound_masses,
                      compound_names):
    network_builder.add_compound(
        mass, name)
for k_f, k_b, s_f, s_b in zip(k_fs, k_bs,
                              s_fs, s_bs):
    network_builder.add_reaction(k_f, k_b,
                                 s_f, s_b)
\end{lstlisting}

Then, the microkinetic model can be integrated by calling the \texttt{integrate} function on the network built by the \texttt{NetworkBuilder} with the start concentrations \texttt{c\_start}, the start time \texttt{t\_start}, initial time step \texttt{dt}, final time \texttt{t\_max}, and integrator choice:

\begin{lstlisting}[language=Python, caption=Solving the microkinetic model with KiNetX.]
network = network_builder.generate()
result = kx.integrate(
    network, c_start,
    t_start, dt,
    kx.Integrator.cash_karp_5,
    integrateByTime=True,
    maxTime=t_max)
\end{lstlisting}
As a result, the integrate function returns the final and maximum concentrations, the integrated absolute concentration fluxes\cite{Bensberg2023} through the compounds and reactions.

\section{Example Application}
\label{sec:example}

\begin{figure}[H]
\centering
\includegraphics[width=0.95\columnwidth]{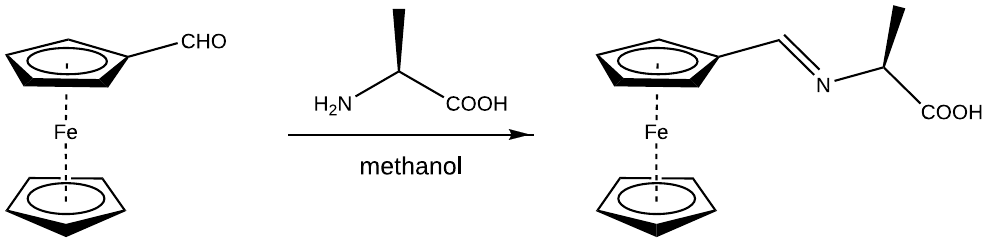}
\caption{Example reaction studied in this work.}
\label{fig:example}
\end{figure}

To illustrate how individual SCINE modules can be combined to perform a complex task, we study the reaction between ferrocene carboxaldehyde and (\textit{S})-alanine (\textit{cf.}, Fig.~\ref{fig:example}). This reaction is the first step of the synthesis of (\textit{R})-$\alpha$-methyl phenylalanine presented by Alonso and Davies in 1995\cite{Alonso1995}. Note that the original reaction sequence involves the sodium salt of (\textit{S})-alanine rather than (\textit{S})-alanine.

All scripts used to set up and carry out this exploration as well as all data collected are available on Zenodo\cite{data_set}. For the sake of simplicity and speed, we conducted the exploration with the GFN2-xTB method\cite{Bannwarth2019}. Furthermore, we did not consider the conformational space of the individual molecules, instead starting from predefined conformations for the two reactants.

To begin the exploration, the code presented in Listing~\ref{code:start} establishes a connection to a database, and stores the structures contained in the files \texttt{ferrocene\_carboxaldehyde.xyz} and \texttt{alanine.xyz} as guess structures in the database. They are then subsequently optimized with the GFN2-xTB method (as defined by the \texttt{model}); these optimizations are done by a Puffin instance, which itself relies on ReaDuct to carry out the actual optimization (ReaDuct, in turn, will rely on an optimizing algorithm implemented in the Utilities package, and the external xtb program to calculate electronic energies, nuclear gradients, and Hessian matrices).

The \texttt{compound\_engine} will automatically analyze the optimized structures, requesting bond orders to be calculated. This information is then processed by Molassembler to create a molecular graph, which is utilized to assign each structure to a compound as explained above. The \texttt{thermo\_engine} automatically requests Hessian calculations for any structures (if such calculations do not yet exist) for later thermochemical analyses. In our minimal example, we use a very simple \texttt{kinetics\_gear} which activates any compound for further exploration. In a more sophisticated approach, one could instead employ a different gear relying on KiNetX to enable the exploration of compounds based on their concentration fluxes (see above).

The \texttt{reaction\_engine} assigns any elementary steps to the correct reaction. At this stage of the exploration, this engine does not alter the database, since no elementary steps are yet present. Finally, the \texttt{scheduling\_engine} takes care of scheduling the individual calculations.

\begin{lstlisting}[language=Python, label=code:start, caption={Setting up the database for the example exploration. For the sake of brevity, only parts of the full initialization script are shown here. The full script is available on Zenodo\cite{data_set}.}]
# Connect to database
manager = db.Manager()
db_name = "scine_demonstration"
ip = '127.0.0.1'
port = '27017'
credentials = db.Credentials(ip, int(port), db_name)
manager.set_credentials(credentials)
manager.connect()

# Define model for all calculations
model = db.Model("gfn2", "gfn2", "")

# Initialize database
wipe = True
if wipe:
    manager.wipe()
    manager.init()
    time.sleep(1.0)
    # Load initial data
    ferrocene_carboxaldehyde = "ferrocene_carboxaldehyde.xyz"
    alanine = "alanine.xyz"

    structure, calculation = insert_initial_structure(
        manager,
        ferrocene_carboxaldehyde,
        0,
        1,
        model,
        settings=default_opt_settings()
    )
    structure, calculation = insert_initial_structure(
        manager,
        alanine,
        0,
        1,
        model,
        settings=default_opt_settings()
    )

# Compound generation and sorting
compound_gear = BasicAggregateHousekeeping()
compound_gear.options.model = model
compound_engine = Engine(credentials)
compound_engine.set_gear(compound_gear)
compound_engine.run()

# Thermo-chemical data completion
thermo_gear = BasicThermoDataCompletion()
thermo_gear.options.model = model
thermo_gear.options.job = db.Job("scine_hessian")
thermo_engine = Engine(credentials)
thermo_engine.set_gear(thermo_gear)
thermo_engine.run()

# Sorting elementary steps into reactions
reaction_gear = BasicReactionHousekeeping()
reaction_engine = Engine(credentials)
reaction_engine.set_gear(reaction_gear)
reaction_engine.run()

# Driving exploration based on kinetics
kinetics_gear = MinimalConnectivityKinetics()  # activate all compounds
kinetics_engine = Engine(credentials)
kinetics_engine.set_gear(kinetics_gear)
kinetics_engine.run()

# Calculation scheduling
scheduling_gear = Scheduler()
scheduling_gear.options.job_counts = {
    "scine_single_point": 500,
    "scine_geometry_optimization": 500,
    "scine_ts_optimization": 500,
    "scine_bond_orders": 500,
    "scine_hessian": 200,
    "scine_react_complex_nt2": 100,
    "scine_dissociation_cut": 100,
    "conformers": 20,
    "final_conformer_deduplication": 20,
    "graph": 1000,
}
scheduling_engine = Engine(credentials)
scheduling_engine.set_gear(scheduling_gear)
scheduling_engine.run()
\end{lstlisting}

After having initialized the database with the two reactants, we can set up reaction trials in a next step. A brute-force approach would result in over 2'000'000 trial calculations; however, we can cut down this number significantly by taking advantage of chemical intuition as shown in Listing~\ref{code:step1}. First, we can focus exclusively on bimolecular reactions between (\textit{S})-alanine and ferrocene carboxaldehyde, leaving away any unimolecular reactions with the setting \texttt{enable\_unimolecular\_trials = False}. Furthermore, we employ the predefined \texttt{CatalystFilter} to restrict bimolecular reaction trials to involve ferrocene carboxaldehyde and (\textit{S})-alanine, excluding trials with two identical molecules (as the name suggests, this filter is particularly useful in catalysis applications, but it can be very handy also in other situations, such as the one we are dealing with here). Finally we can restrict the exploration to selected functional groups with the variable \texttt{reactive\_site\_filter}. In our example, we restrict the reactivity to the aldehyde and the amino groups.

\begin{lstlisting}[language=Python, label=code:step1, caption={Setting up reaction trials between alanine and ferrocene carboxaldehyde. For the sake of brevity, only parts of the full initialization script are shown here. The full script is available on Zenodo\cite{data_set}.}]
nt_job = db.Job("scine_react_complex_nt2")
nt_settings = default_nt_settings()

elementary_step_gear = BruteForceElementarySteps()
elementary_step_gear.trial_generator = BondBased()
elementary_step_gear.trial_generator.options.model = model
elementary_step_gear.options.enable_bimolecular_trials = True
elementary_step_gear.options.enable_unimolecular_trials = False

# Add filters
elementary_step_gear.trial_generator.reactive_site_filter = AtomRuleBasedFilter(DistanceRuleSet({
    'C': FunctionalGroupRule(0, 'C', (3, 3), {'O': 1, 'H': 1, 'C': 1}, True),
    'O': DistanceRuleAndArray([FunctionalGroupRule(0, 'O', (1, 1), {'C': 1}, True), FunctionalGroupRule(1, 'C', (3, 3), {'O': 1, 'C': 1, 'H': 1}, True)]),
    'N': FunctionalGroupRule(0, 'N', (3, 3), {'C': 1, 'H': 2}, True),
    'H': FunctionalGroupRule(1, 'N', (3, 3), {'H': 2, 'C': 1}, True)}))

elementary_step_gear.aggregate_filter = CatalystFilter({'Fe': 1})

# Run
elementary_step_engine = Engine(credentials)
elementary_step_engine.set_gear(elementary_step_gear)
elementary_step_engine.run()
\end{lstlisting}

\begin{figure}[H]
\centering
\includegraphics[width=0.95\columnwidth]{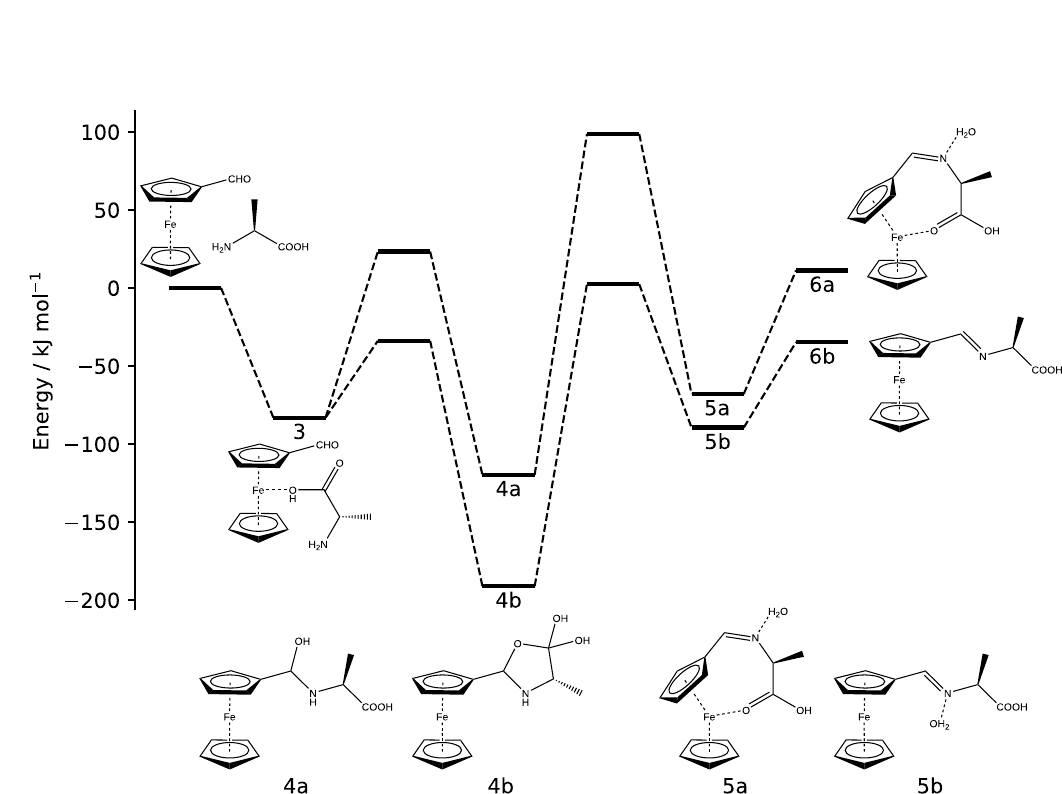}
\caption{Energy diagram of the example reaction.}
\label{fig:energy_diagram}
\end{figure}

These reaction trials are subsequently evaluated by Puffin as depicted in Fig.~\ref{fig:reaction_trial_schema}. Several elementary steps are found in which the nitrogen atom attacks the carbonyl carbon atom of ferrocene carboxaldehyde, resulting in intermediates 4a and 4b in Fig.~\ref{fig:energy_diagram}. Note that before this attack, (\textit{S})-alanine coordinates to the iron atom (intermediate 3 in Fig.~\ref{fig:energy_diagram})\,---\,Chemoton finds this reaction even though the metal center was not included as a reactive site when setting up the reaction trials.

Starting from intermediates 4a and 4b, new reaction trials were created with slightly different filter settings (the script to create these additional trials is also available on Zenodo\cite{data_set}). These result in several elementary steps eliminating water, leading to intermediates 5a and 5b (\textit{cf.}, Fig.~\ref{fig:energy_diagram}). Directly after the formation of water, it forms a hydrogen bond to the nitrogen atom; breaking this hydrogen bond yields the final product (6a and 6b in Fig.~\ref{fig:energy_diagram}). Note that Chemoton does not only find the product 6b as shown in Fig.~\ref{fig:example}, but also the compound 6a in which the carboxylic acid coordinates to the iron atom. This necessitates a deformation of the ferrocene moiety, significantly tilting one of the cyclopentadienyl rings, which results in 6a being less stable than 6b. 

\section{Conclusions}
\label{sec:conclusion}
Here, we presented the modules of the SCINE project that provide a framework for quantum chemical calculations on isolated and embedded molecular structures that are to be arranged in some conceptual framework. Typical frameworks that provide contextual dependence for these structures are chemical reaction networks or high-throughput virtual screening campaigns in an attempt to optimize or design certain properties (such as linear or nonlinear optical properties or specific reactivity patterns). Although the focus of our work has been mostly on the former application area so far, the latter is easily realizable within the SCINE framework. Modules that produce data on molecular structures can be easily combined to accomplish high-level tasks (as highlighted by the steering through a reaction mechanism during reaction network exploration with the Steering Wheel\cite{Steiner2024}). Data is stored in and recovered from a central database structure. The Molassembler module produces unique graph information for the identification of all kinds of molecular structure, which is key for the identification, storage, and exploitation of Cartesian coordinates as molecules. The high degree of modularity and interoperability of the SCINE programs has facilitated the development of the cloud-based reaction mechanism screening workflow AutoRXN\cite{Unsleber2023a}. In general, the SCINE project has established a platform of high usability for further developments in the field of high-throughput computational campaigns.

\begin{acknowledgments}

This publication was created as part of NCCR Catalysis, a National Centre of Competence in Research funded by the Swiss National Science Foundation (grant number 180544). 
M.~E.~has been supported by an ETH Zurich Postdoctoral Fellowship and M.~S.~by a Swiss Government Excellence Scholarship for Foreign Scholars and Artists.
We gratefully acknowledge financial support for M.~M.~through ETH Research Grant ETH-43 20-2.

We would like to thank Christoph Brunken for his foundational work on the Swoose module. Moreover, we are grateful to other members of the Reiher research group, students, and collaborators who have contributed to parts of the software (see the Zenodo references cited in this work). In particular, we would like to mention
Alberto Baiardi,    
Robin Feldmann,
Nina Glaser, 
Stefan Gugler,
Moritz Haag, 
Michael Heuer,
Tamara Husch, 
Lucas Lang,
Severin Polonius,
Jonny Proppe,
Gregor Simm,
Christopher Stein, 
and
Alain Vaucher.
\end{acknowledgments}

\section*{Data Availability Statement}

The data that support the findings of this study are openly available at Zenodo at http://doi.org/10.5281/zenodo.11048378.
The source code of the SCINE software modules is available on \url{https://github.com/qcscine} as well as on Zenodo\cite{chemoton310, molassembler201, puffin130, swoose100, kinetx200, readuct510, sparrow500, autocas210, heron100, core600, utilities900, database130}. Additional information about the SCINE modules can be obtained on \url{https://scine.ethz.ch}.

\bibliography{references}

\end{document}